\documentclass{aa}
\usepackage[varg]{txfonts}
\usepackage{natbib}
\usepackage{xcolor}
\usepackage{braket}
\usepackage{amsmath}

\newcommand{\B}[1]{\boldsymbol{#1}}
\newcommand{\R}[1]{\mathrm{#1}}
\newcommand{\red}[1]{\textcolor{red}{#1}}

\begin{document}

\title{Regularities in the spectrum of chaotic p-modes in rapidly rotating stars}

\author{Benjamin Evano\inst{1,2}
\and François Ligni\`eres\inst{1}
\and Bertrand Georgeot\inst{2}}

% \offprints{R. Plemmons, \email{plemmons@...}}

\institute{Institut de Recherche en Astrophysique et Plan\'etologie, Universit\'e de Toulouse, CNRS, CNES, UPS, France
  \and Laboratoire de Physique Th\'eorique, IRSAMC, Universit\'e de Toulouse, CNRS, UPS, France}
  
\date{}

\abstract {Interpreting the oscillations of massive and intermediate mass stars remains a challenging task. In fast rotators, the oscillation spectrum of p-modes is a superposition of sub-spectra which corresponds to different types of modes, among which island modes and chaotic modes are expected to be the most visible. This paper is focused on chaotic modes, which have not been thoroughly studied before.} {We study the properties of high frequency chaotic p-modes in a polytropic model. Unexpected peaks appear in the frequency autocorrelations of the spectra. Our goal is to find a physical interpretation for these peaks and also to provide an overview of the mode properties.} {We used the 2D oscillation code "TOP" to produce the modes and acoustic ray simulations to explore the wave properties in the asymptotic regime. Using the tools developed in the field of quantum chaos (or wave chaos), we derived an expression for the frequency autocorrelation involving the travel time of acoustic rays.} {Chaotic mode spectra were previously thought to be irregular, that is, described only through their statistical properties. Our analysis shows the existence, in chaotic mode spectra, of a pseudo large separation. This means that chaotic modes are organized in series, such that the modes in each series follow a nearly regular frequency spacing. The pseudo large separation of chaotic modes is very close to the large separation of island modes. Its value is related to the sound speed averaged over the meridional plane of the star. In addition to the pseudo large separation, other correlations appear in the numerically calculated spectra. We explain their origin by the trapping of acoustic rays near the stable islands.}{}

\maketitle

\section{Introduction}

Despite the many successful advancements made in stellar seismology, we are still unable to unlock most of the information contained in the acoustic oscillations of typical massive and intermediate mass stars. The observed pulsational behavior of these stars in the range of acoustic frequencies is diverse and has so far resisted a meaningful empirical classification. In addition to evolutionary effects, both the nonlinear processes that set the mode amplitudes and the sensitivity of the spectrum organization to the unknown rotation rate contribute to the observed diversity \citep{Bowman2018}. 

Progress in modeling the rotational effects together with the flow of high quality data from ultra precise space photometry missions, nevertheless, revive the interest for the seismology of massive and intermediate mass stars. An important step was the detection of regular frequency patterns analogous to the solar-like large separation in $\delta$ Scuti stars \citep{hernandez_asteroseismic_2009, Hernandez2013, garcia_hernandez_observational_2015, Paparo2016, Michel2017}. These patterns were predicted from the first oscillation model, that took realistic centrifugal distortion into account \citep{lignieres_acoustic_2006}, up to the more realistic models to date \citep{reese_frequency_2017}. Further progress is expected from TESS and PLATO missions that will include bright stars that can be better constrained through spectroscopy. 

To model the acoustic spectrum of rapidly rotating stars, 2D oscillation codes that take 
full account of the effect of rotation on the oscillations were developed \citep{reese_acoustic_2006, reese_pulsation_2009, ouazzani_pulsations_2012}. They can be run for different models of centrifugally deformed stars, from polytropic models \citep{reese_acoustic_2006} to more sophisticated ones \citep{reese_pulsation_2009, ouazzani_pulsations_2015}. Exploring the acoustic spectrum as a function of the star rotation is not a straightforward process, in particular, because the density of frequencies increases with the numerical resolution and the mode classification can be tedious \citep{Ballot2013}. First attempts were restricted to low spherical-harmonic degree modes, which were followed by progressively increasing the rotation rate \citep{lignieres_acoustic_2006, reese_acoustic_2006, reese_regular_2008, pasek_regular_2012}. More complete spectra have then been obtained at fixed rotation rates \citep{lignieres_asymptotic_2009, reese_pulsation_2009, ouazzani_pulsations_2015, reese_frequency_2017}. An automatic classification of the computed mode using neural network methods has recently been proposed \citep{mirouh_mode_2019}. 

Understanding the spectrum organization is key to constructing seismic tools, the detection of regular patterns in  $\delta$ Scuti stars being a good example. The structure of the spectrum is expected to best reveal itself in the asymptotic regime, that is, at high-frequencies for acoustic modes. This motivates high-frequency computations even if massive and intermediate mass stars oscillate at lower acoustic frequencies. The asymptotic regime is also amenable to theoretical descriptions. In the short wavelength "WKB" limit, acoustic waves are described by rays whose propagation obeys Hamiltonian dynamics. In \citet{lignieres_asymptotic_2009}, the acoustic ray dynamics of rapidly rotating stars was studied and semiclassical concepts and methods developed in quantum physics were used to infer the asymptotic properties of the associated acoustic spectrum. Reflecting the phase space structure, different families of modes were identified. Above some rotation, two families are most likely to be observed : the 2-period island modes and the chaotic modes. The first family was studied in detail through numerical computations \citep{lignieres_acoustic_2006, reese_regular_2008, reese_pulsation_2009, ouazzani_pulsations_2015} and semi-analytical models \citep{pasek_regular_2011, pasek_regular_2012}. These modes show regular frequency spacings and should be the most visible in the observed spectra. They have been cited to explain the frequency patterns observed in some $\delta$ Scuti stars \citep{garcia_hernandez_observational_2015}.

Chaotic modes are called chaotic because they result from the constructive interference of waves which, in the short wavelength limit, have chaotic trajectories. They were first studied in quantum physics, showing distinctive features, such as ergodicity or the universality of the nearest neighbor statistics of their spectrum. \citet{lignieres_asymptotic_2009} identify $\sim 200$ axisymmetric chaotic modes at a given rotation rate and verify that they follow the expected statistics.

In this paper the asymptotic properties of chaotic acoustic modes in rapidly rotating stars is investigated in detail. We construct a large set of high-frequency chaotic modes, computed at various rotation rates and azimuthal numbers $m$, analyze the mode properties, and interpret them using semiclassical methods.

Among these properties, the presence of peaks in the auto-correlation of the chaotic mode spectrum had not been reported in the experimental or modeled wave systems considered in the fields of quantum chaos. This justified a publication dedicated to this particular point in a physics journal \citep{evano_correlations_2019}. The present paper complements and adds to this publication, putting emphasis on the asteroseismic applications.

The paper is organized as follows : we first introduce the formalism and numerical methods, then present the properties of high frequency chaotic modes (Section 3), and develop the semiclassical theory which explains the features observed (Section 4). The Appendix presents detailed derivations needed for Section 4.

\section{Formalism and numerical methods}

In this section, we first introduce the equations and tools used to study propagating pressure waves in stars. Then we present the range of numerically computed high-frequency pressure modes and finally the method used to classify the modes and obtain a set of high-frequency chaotic modes for different rotation rates.

\subsection{Pressure waves and their ray limit in rapidly rotating stars}

We consider adiabatic pressure perturbations in a self-gravitating gas. We are focusing on p-modes in the high-frequency regime, and as is usual in this case we apply the Cowling approximation, neglecting the perturbations of the gravitational potential. This approximation is known to be valid for high-frequency perturbations in non rotating stars \citep{aerts_asteroseismology_2010}. We also neglect the Coriolis force since  its influence on the pulsation frequencies is known to be weak in the high-frequency regime, see  \citet{reese_acoustic_2006}. Finally, we can discard the gravity waves since the Brunt-Väisälä frequency is very small compared to the high p-mode frequencies. With all these approximations taken into account, the adiabatic pressure perturbations obey a Helmholtz-like equation derived in \citet{lignieres_asymptotic_2009} :

\begin{equation}
\label{helmoltz}
    \Delta \Psi + \frac{\omega^2 - \omega_c^2}{c_s^2} \Psi = 0,
\end{equation}

\noindent where $\Psi = \hat{P}/f$, with $\hat{P}$ the complex amplitude associated with the time-harmonic pressure perturbation $P = \R{Re}[\hat{P} \, \exp(-i \omega t)]$, $f$ a function of the background model, $\omega_c$ is a cut-off frequency of the model and $c_s$ is the inhomogeneous sound velocity. \\

The short-wavelength approximation of Eq.~\ref{helmoltz} leads to the eikonal equation

\begin{equation}
\label{eq:eikonal}
\omega^2 = c_s^2 k^2 + \omega_c^2.
\end{equation}

\noindent This equation can be put in the form of a Hamiltonian system describing the propagation of acoustic rays \citep{ott_chaos_1993}, by setting $H = \omega = (c_s^2 k^2 + \omega_c^2)^{1/2}$. The motion takes place in a meridional plane rotating with the ray at an angular velocity $\R{d}\phi / \R{d}t = \tilde{L}_z / \R{d}^2$, where $\tilde{L}_z = r \sin \theta \, k_{\phi}/\omega$ is a constant of motion and $d$ is the distance to the axis of rotation. To compute the ray paths we use an alternative Hamiltonian form derived in \citet{lignieres_asymptotic_2009} as :

\begin{align}
    H' &= \frac{\B{\tilde{k}^2_p}}{2} + W, \\
    W &= - \frac{1}{2 c_s^2} \left(1-\frac{\omega_c^2}{\omega^2}\right) + \frac{\tilde{L}_z^2}{2d^2},
\end{align}

\noindent where $\B{\tilde{k}_p}$ is the frequency-scaled wave vector $\B{\tilde{k}} = \B{k}/\omega$ projected onto the meridional plane and $W$ is the potential. The corresponding dynamical system is

\begin{equation}
\frac{\R{d} \B{x}}{\R{d}t} = \B{\tilde{k}_p}, \quad \frac{\R{d} \B{\tilde{k}_p}}{\R{d}t} = - \B{\nabla} W,
\end{equation}

\noindent where $\R{d}t = c_s \R{d}s/ \left(1-\frac{\omega_c^2}{\omega^2}\right)^{1/2}$, is a time-like variable, $s$ being the curvilinear coordinate along the ray. This system is then integrated using a fifth order Runge-Kutta method. To simplify the notation, we now refer to $\B{\tilde{k}_p}$ as simply $\B{\tilde{k}}$.\\

\begin{figure}[!htbp]
\centerline{\includegraphics[width=0.5\textwidth]{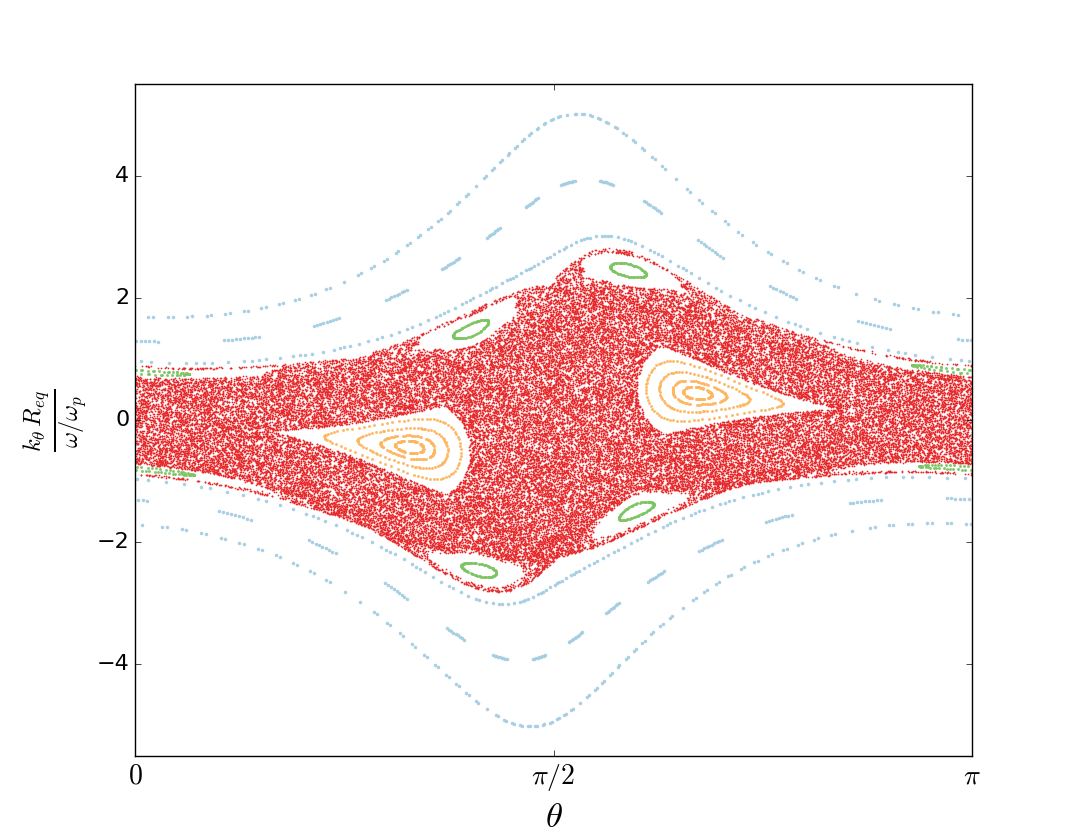}}
\caption{PSS at rotation rate $\Omega / \Omega_k = 0.589$ with $\tilde{L}_z = 0$, where $\theta$ is the colatitude and $k_{\theta}$ the projection of the wave vector on the line tangent to the $r_p(\theta) = r_s(\theta) - z$ curve. Two stable islands are embedded in a chaotic region, which is itself surrounded by whispering gallery rays. The different types of rays are color coded, blue : whispering gallery, green : 6-period island, yellow : 2-period island and red : ergodic.}
\label{PSS_59}
\end{figure}

The motion of acoustic rays inside the meridional plane corresponds to a dynamical system with $N_d=2$ degrees of freedom. Because the wave frequency is a conserved quantity, the phase space trajectories evolve on a $2N_d-1$ surface. To reveal the properties of a dynamical system, it is customary to define a surface of lower dimension, called a Poincar\'e Surface of section (PSS), by fixing an additional parameter \citep{gutzwiller_chaos_1990, ott_chaos_1993}. There are multiple possible choices of PSS, the recommendations for a faithful representation are given in \citep{ott_chaos_1993}. We have chosen a surface at constant radial distance $z$ from the star surface $r_s(\theta)$ : $r_{p}(\theta) = r_s(\theta) - z$, where $r_{p}(\theta)$ is the radial coordinate of the PSS. An example of such a PSS is shown in Fig.~\ref{PSS_59}. Each dot on this figure corresponds to the crossing of the PSS with a ray approaching the surface. Chaotic zones correspond to trajectories filling densely an area, whereas regular trajectories are constrained to remain on a closed curve. The chaotic dynamics implies ergodicity in a certain phase space zone, and exponential separation of nearby trajectories, whose rate is measured by the Lyapunov exponents \citep{ott_chaos_1993}.

\begin{figure}[!htbp]
\centerline{\includegraphics[width=0.5\textwidth]{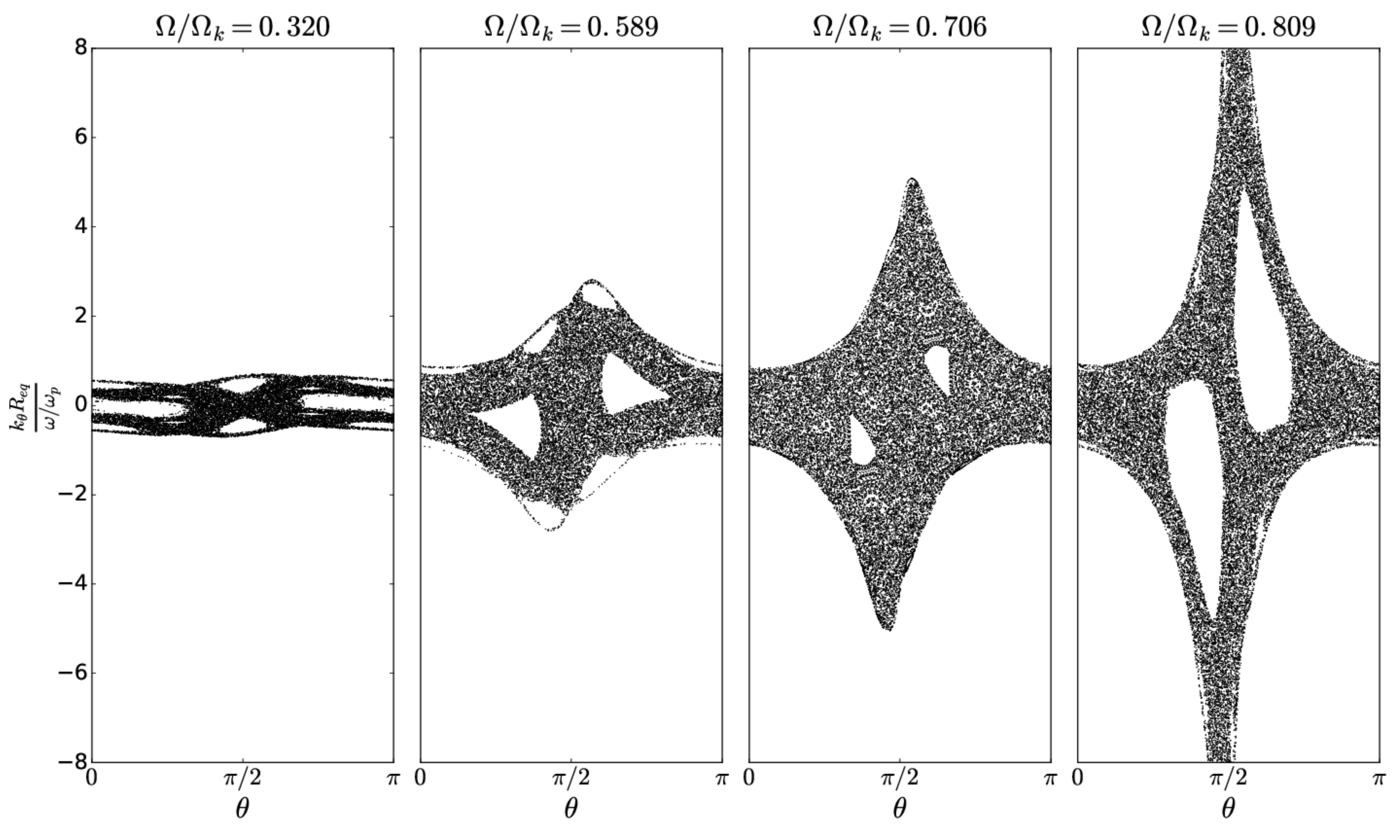}}
\caption{Chaotic zone of the PSS at increasing values of the rotation rate. The chaotic zone grows monotonically, which is not the case for the main island zone.}
\label{PSS_evolution}
\end{figure}

\noindent The system undergoes a KAM-like transition (named from a famous theorem from Kolmogorov, Arnold and Moser describing the typical transition to chaos, see for example \citet{ott_chaos_1993}), with the rotation rate $\Omega$ playing the role of the perturbation. It is known that in this case the phase space contains chaotic zones, island orbits around stable periodic orbits and KAM tori reminiscent of the $\Omega = 0$ system. This is represented in Fig.~\ref{PSS_59}. We did not draw the small island structures in the domain where $|k_{\theta}/\omega|$ is high. Ergodic trajectories appear as soon as the spherical symmetry is broken. At low rotation rates the chaotic zone is very small. However, it expands considerably as the star is flattened by rotation. The rotation rate of the model of star will be given in terms of the Keplerian break-up rotation rate $\Omega_k = (G M / R_{eq}^3)^{1/2}$, where $G$ is Newton's constant, $M$ is the mass of the star and $R_{eq}$ the equatorial radius. Near $\Omega / \Omega_k \simeq 0.32$ the chaotic zone is comparable in size to the main stable island and continues to grow afterward (see Fig.~\ref{PSS_evolution}). The ratio $\mathcal{V}_{\R{I}} / \mathcal{V}_{\R{II}}$, where $\mathcal{V}_{\R{I}}$ is the volume occupied by the main stable island and $\mathcal{V}_{\R{II}}$ is the volume of the main chaotic zone, is a relevant quantity as it indicates (as we shall see) the proportion of island modes to chaotic modes. Measuring the areas $\mathcal{A}_{\R{I}}$ and $\mathcal{A}_{\R{II}}$ on the PSS, corresponding respectively to the main island and chaotic zones, we found out that the function $\mathcal{A}_{\R{I}}(\Omega) / \mathcal{A}_{\R{II}}(\Omega)$ is not monotonic. In fact, the ratio of the 2-period island over the chaotic zone shrinks beyond $\Omega / \Omega_k = 0.658$, reaches a minimum around $\Omega / \Omega_k = 0.682$, then increases again from this point onward. This non-monotonic behavior was not detected in \citet{lignieres_asymptotic_2009} as rotation rates around $\Omega / \Omega_k = 0.682$ were not considered. \\

\begin{figure}[!htbp]
\centerline{\includegraphics[width=0.5\textwidth]{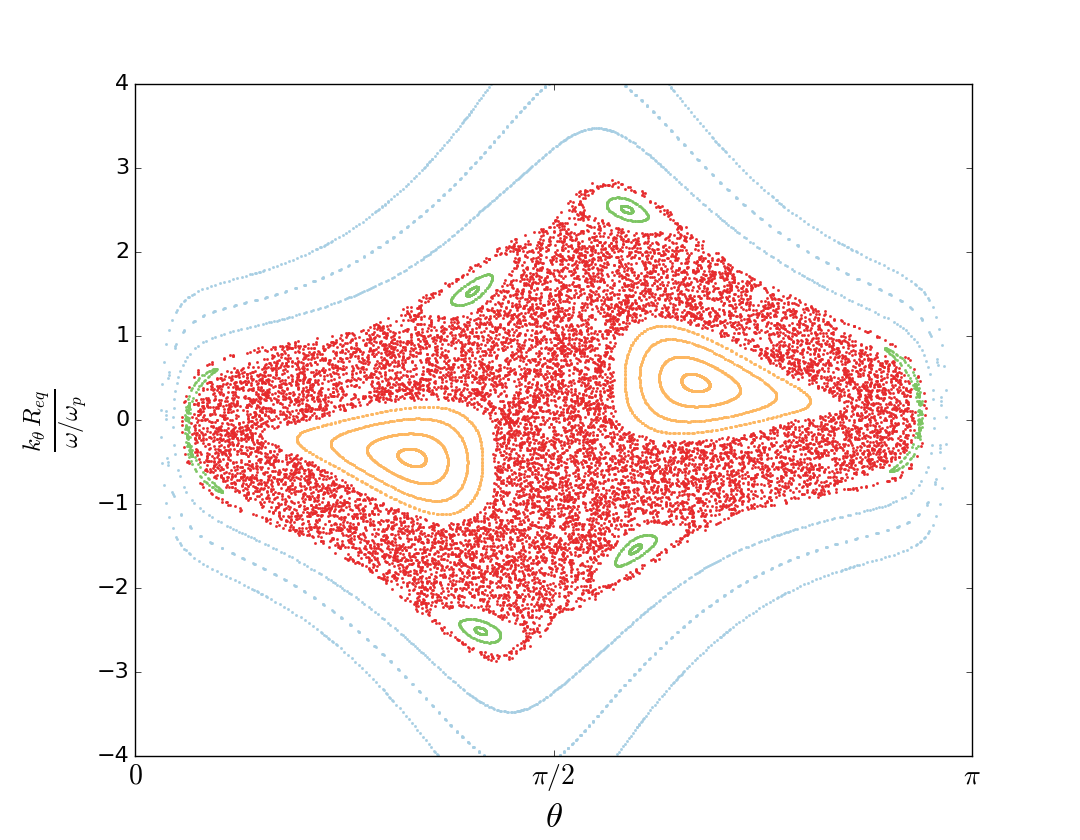}}
\caption{PSS at $\Omega / \Omega_k = 0.589$ with $\tilde{L}_z = 0.16 / \omega_p$, where $\theta$ is the colatitude and $k_{\theta}$ the projection of the wave vector on the line tangent to the $r_p(\theta) = r_s(\theta) - z$ curve. The phase space structures are similar to those presented in Fig.~\ref{PSS_59}}
\label{PSS_59_m4}
\end{figure}

The non axisymmetric ray dynamics $\tilde{L}_z \neq 0$, shown in Fig.~\ref{PSS_59_m4}, is qualitatively similar, but the expansion of the chaotic zone is retarded. Moreover, the domain of propagation is reduced. Indeed, the ray has azimuthal velocity and, as a result, cannot come arbitrarily close to the rotation axis. In the meridional plane, this phenomenon translates into a condition on the colatitude \citep{lignieres_asymptotic_2009} $- \arcsin(|\tilde{L}_z|/\tilde{L}) < \theta < \arcsin(|\tilde{L}_z|/\tilde{L})$, $\tilde{L}$ being the frequency-scaled angular momentum norm. \\

The semiclassical theory of such a mixed phase space with both regular and chaotic zones, due to Berry and Robnik \citep{berry_semiclassical_1984}, indicates that regular and chaotic modes can be associated to the different phase space areas and form independent subspectra . This prediction was verified in \citet{lignieres_asymptotic_2009} by computing high-frequency modes and classifying them according to the phase space structure. 

\subsection{The set of numerically computed high-frequency p-modes}

\begin{figure}[!htbp]
\centerline{\includegraphics[width=1\columnwidth]{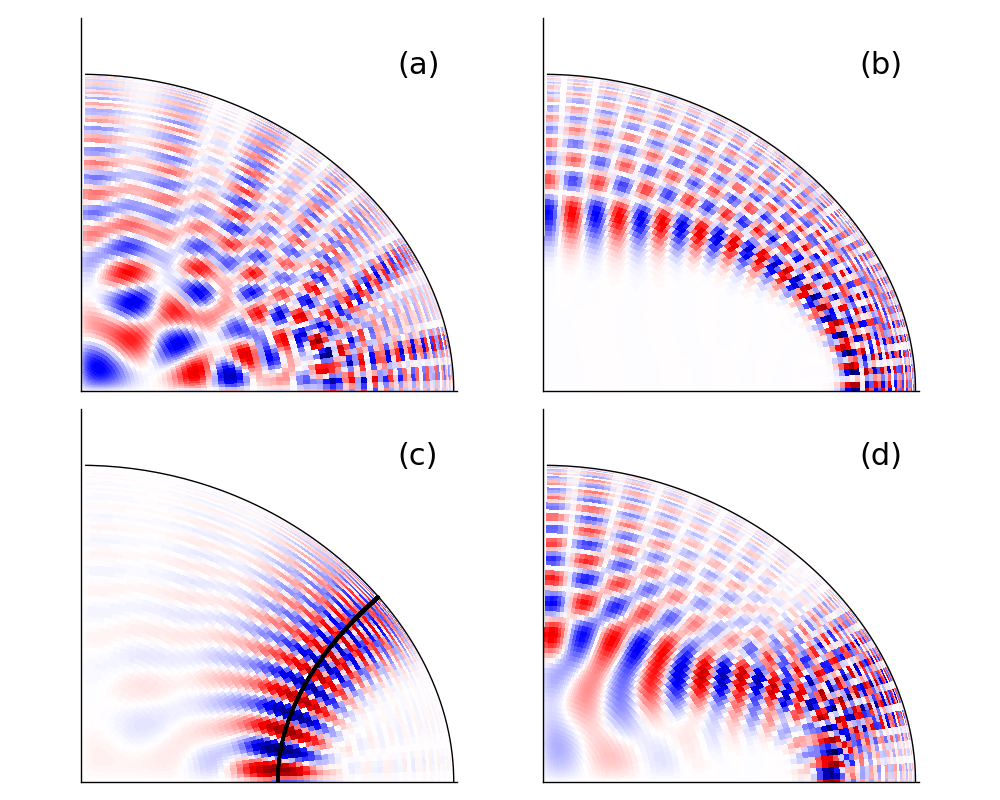}}
\caption{Four odd axisymmetric modes at rotation $\Omega / \Omega_k = 0.589$ : (a) chaotic mode, (b) whispering gallery mode, (c) 2-period island mode ($\ell = 0$), with a black line indicating the central periodic orbit, and (d) 6-period island mode. The figure shows the scaled pressure amplitude $\Psi =  P \sqrt{d/\rho_0}$, with $d$ the distance to the rotation axis and $\rho_0$ the equilibrium density.}
\label{4_modes}
\end{figure}

As for the ray dynamics, the star is modeled by a uniformly rotating, self-gravitating monoatomic perfect gaz of adiabatic exponent $\Gamma = 5/3$. We impose that the pressure and density satisfy a polytropic relation $P \propto \rho^{1+1/\mu}$, with $\mu =3$ \citep{hansen_stellar_2004}. As we are interested in high-frequency, and thus small-wavelength, p-modes, attention has to be paid to numerical resolution. The stellar model is calculated using spectral methods with Chebyshev polynomials in a pseudo radial direction and Legendre polynomial in latitude, corresponding to $N_j = 96, 128, 140$ radial points and $N_t \leq 185$ latitudinal points. To compute its oscillation modes, we use the code TOP described in \citep{reese_acoustic_2006, reese_pulsation_2009}. Modes are computed using $N_j$ Chebyshev polynomials $T_j$ and $N_l$ spherical harmonics $Y_l^m$, through the decomposition :

\begin{equation}
    \Psi(r, \theta, \phi) = \sum_{l=0}^{N_l} \, \left(\sum_{j=0}^{N_j} a_j ^{l, m} T_j(2r-1)\right), \, Y_l^m(\theta, \phi),
\label{mode_decomposition}
\end{equation}

\noindent where the degrees $l$ are either odd or even integers \citep{reese_acoustic_2006}. The needed resolution is determined by the smallest scale on which the mode amplitude varies. Physically, it is directly related to the highest values taken by the components of the wave vector $k_{\theta}$ and $k_r$. Using the PSS of a given stellar model, we can determine the upper limit of the wave vector components $\tilde{k}_{\theta}^{max}$ or $\tilde{k}_r^{max}$ associated with a particular phase space structure. The chaotic zone growth seen in Fig.~\ref{PSS_evolution} indicates that the required angular resolution to compute chaotic modes increases rapidly with rotation. Therefore, computing high-frequency chaotic modes is more and more demanding as the model approaches $\Omega_k$. The central chaotic zone is bounded by whispering gallery rays in the full phase space. Thus we know that, in a given frequency domain, the required resolution to compute all chaotic modes is attained if we are able to produce a few well resolved whispering gallery modes. With these constraints, we produced spectra at six different rotations : $\Omega / \Omega_k = 0.481, 0.545, 0.589, 0.658$, $0.706$ and $0.809$ with frequencies ranging from $\sim 23 \, \omega_p$ to $\sim 47 \, \omega_p$, with $\omega_p = (GM/R_p^3)^{1/2}$, where $R_p$ is the polar radius. A typical example of each kind of mode is shown in Fig.~\ref{4_modes} : a whispering gallery mode corresponding to a KAM torus, a 6-period island mode corresponding to the 6-period island chain, a 2-period island mode corresponding to the 2-period island chain and a chaotic mode corresponding to the chaotic zone. To construct a large set of chaotic modes requires to identify them among all the computed modes.

\subsection{Mode identification}
\label{sec:mode_identification}

In this section we explain our methodology to isolate chaotic modes. To achieve this goal we proceed by elimination. The basic idea is to identify all regular modes and remove them from the dataset until only chaotic modes remain. \\

To begin with, we identify 2-period island modes using the fact that their frequency spectrum is regular and of the form $\omega_{n,l,m} = n \delta_n(m) + \ell \delta_{\ell}(m) + \alpha$, where $n$ (resp. $\ell$) is the number of nodes along (resp. transverse to) the central orbit $\gamma$ of the island, shown in Fig.~\ref{4_modes}, and $\alpha$ is a constant \citep{pasek_regular_2012}. Thus the frequency spectrum is completely determined by the regular spacings $\delta_n$ and $\delta_\ell$, with the half large separation $\delta_n(m) = 2 \pi / \oint_{\gamma} (\R{d}s/\tilde{c}_s)$ \footnote{The large separation is expressed as $\Delta_i = 2 \pi / \int_{\gamma} (\R{d}s/\tilde{c}_s)$, where the integral is calculated along the path $\gamma$ between two points on the surface, which is half the full period of the orbit.}. However, the regularity of the 2-period island modes may be altered by two aspects : first the deviations from asymptotic theory, expected at finite frequency, and secondly the occurrence of avoided crossings between an island mode and a chaotic mode. In practice, the island mode spacing is rigid enough to identify them. In the case of an avoided crossing, we choose arbitrarily one of the two modes as being an island mode and discard the other one. The impact of this choice on the spectrum is weak since the frequencies of two modes in the process of an avoided crossing are very close. In the same way, 6-period island modes are identified on the basis of their regularity \citep{lignieres_asymptotic_2009}. \\

\begin{figure}[!htbp]
\centerline{\includegraphics[width=1\columnwidth]{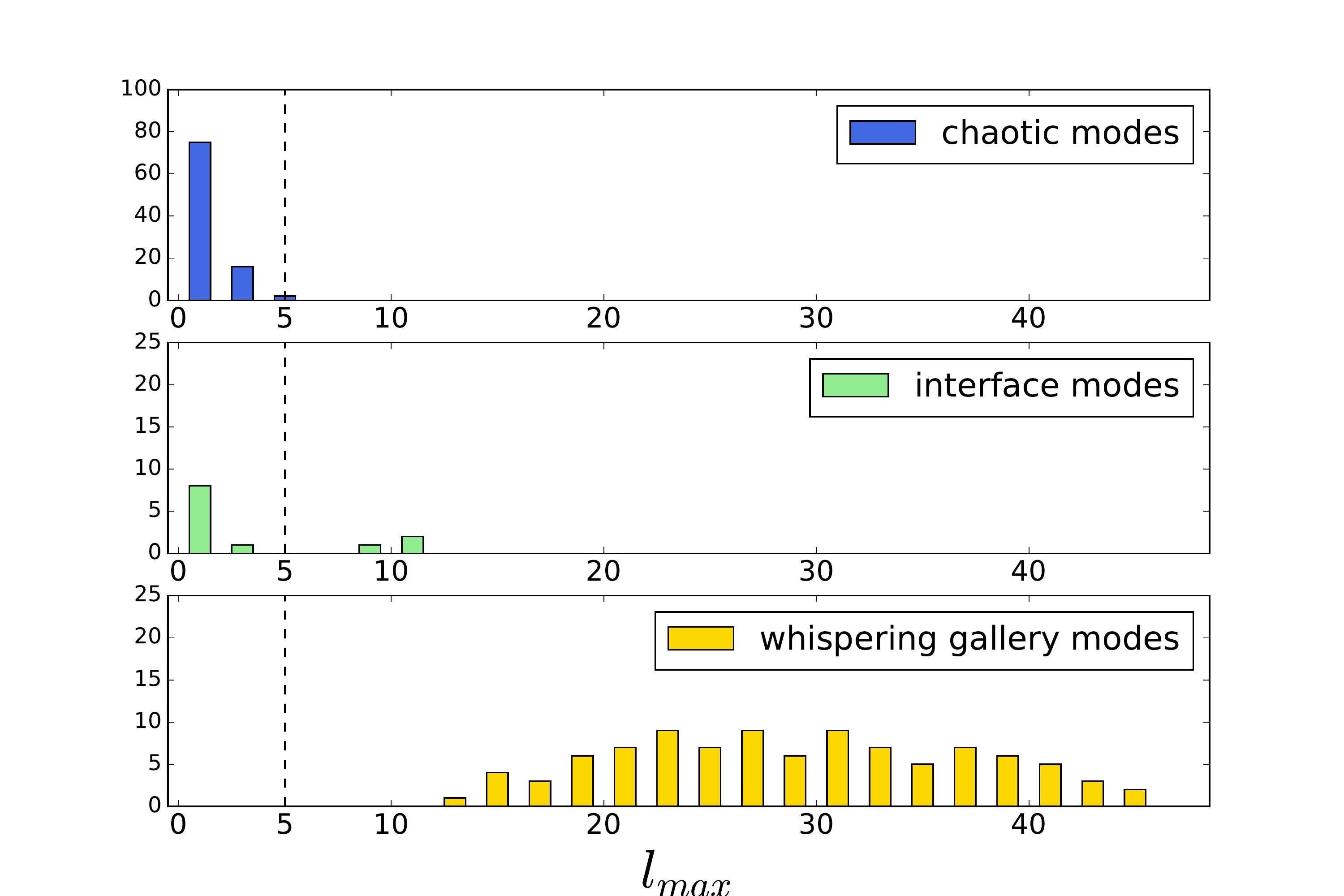}}
\caption{Histograms showing the distribution of three types of odd modes as a function of their dominant degree $l_{max}$, at $\Omega/\Omega_k = 0.589$ in the frequency domain $25.6 \, \omega_p$ to $33.5 \, \omega_p$. The existence of a gap between the highest value of $l_{max}$ for chaotic modes, marked by a dashed line, and the lowest value of $l_{max}$ for whispering gallery modes is used to automatically separate these two types of modes.}
\label{lmax}
\end{figure}

\noindent Then we remove the whispering gallery modes automatically. The PSS shows that the $\tilde{k}_{\theta}^{max}$ of any whispering gallery ray is higher than the $\tilde{k}_{\theta}^{max}$ of the chaotic zone. This means that whispering gallery modes vary on smaller latitudinal scales than chaotic modes, meaning that they have more angular nodes. Therefore, we expect their spherical harmonic expansion, Eq.~\ref{mode_decomposition}, to be dominated by the high degree components.

\noindent For a given $m$, we sum over $j$ to get the averaged coefficients $\bar{a}^l \equiv \tfrac{1}{N_j} \sum_{j=0}^{N_j} |a_j^l|$. Thus the dominant degree $l_{max}$ corresponds to the value of $l$ where $\bar{a}^l$ is the greatest. In the bottom panel of Fig.~\ref{lmax}, the distribution of whispering gallery modes with respect to $l_{max}$ is shown for a given rotation. There is indeed no whispering gallery mode below $l_{max} = 10$. The top panel displays the same distribution for chaotic modes, showing that they are instead dominated by their low degree components. In fact, there is a critical value $l_c$ of $l_{max}$ such that $l_{max} \le l_c$ for every chaotic mode and $l_{max} > l_c$ for any whispering gallery mode. Therefore, it is sufficient to find $l_c$ to remove all whispering gallery modes from the dataset.

\begin{figure}[!htbp]
\centerline{\includegraphics[width=1\columnwidth]{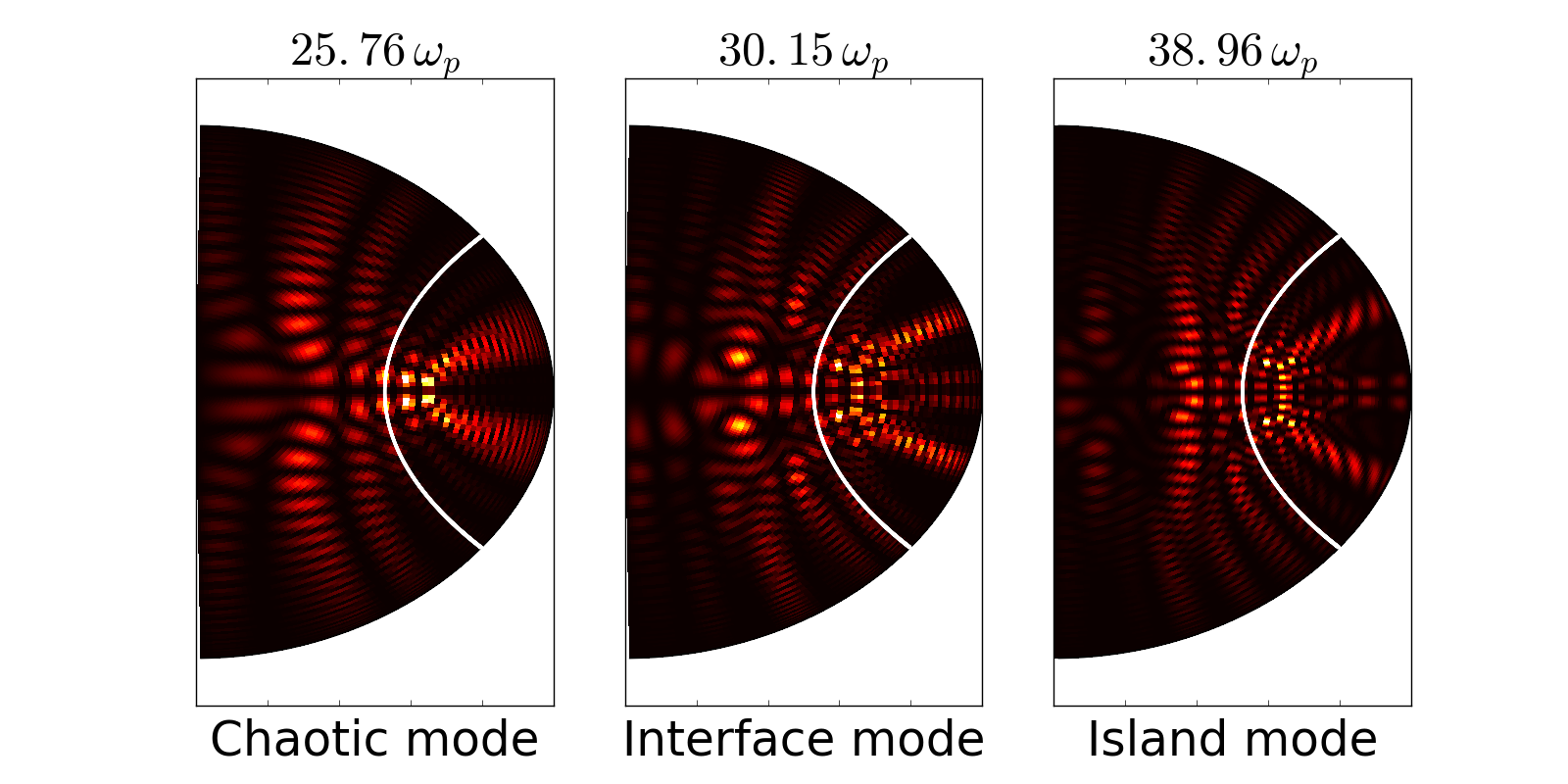}}
\caption{Three modes at $\Omega / \Omega_k = 0.589$ belonging to series 1. From left to right, we see a chaotic mode, an interface mode and an island mode with $\ell = 4$. The thick white line is the main island central orbit. The modes intensity $|\Psi|^2$ is represented, where $\Psi$ is the scaled pressure amplitude as in Fig.~\ref{4_modes}.}
\label{island_interface}
\end{figure}

\noindent  At finite frequency, here $\omega \le 47 \omega_p$, the Berry-Robnik regime is not perfectly satisfied \citep{vidmar2007beyond}. Therefore we expect that some modes cannot be classified in a unique category. We find such modes and call them interface modes. There are two types of interface modes : whispering gallery-like and island-like. An example of island-like interface mode is presented in the central panel of Fig.~\ref{island_interface}. By nature, interface modes have properties of both chaotic modes and either island or whispering gallery modes. To distinguish between "well defined" modes and interface modes, we can look either at their amplitude pattern or at their Husimi distribution, which is a projection of the eigenmode in phase space \citep{chang_evolution_1986, lignieres_asymptotic_2009}. The distinction between chaotic modes and interface modes is not clear-cut, whether we use the amplitude of the modes or the Husimi distributions as a guide. Any time we encountered an ambivalent mode, we applied the following rules to categorize it. If the amplitude is high near an internal caustic, then the mode is a whispering gallery-like interface mode. The difference between chaotic, island-like and island modes is illustrated in Fig~\ref{island_interface}. We consider that a mode is not chaotic if its amplitude is concentrated around the central orbit of the main island. In parallel, we examine the placement of the mode in the spectrum, and look for regularities. Even at the highest frequencies considered, $\ell=4$ island modes cannot be distinguished from interface modes without ambiguity from the amplitudes. \\

\begin{figure}[!htbp]
\centerline{\includegraphics[width=1\columnwidth]{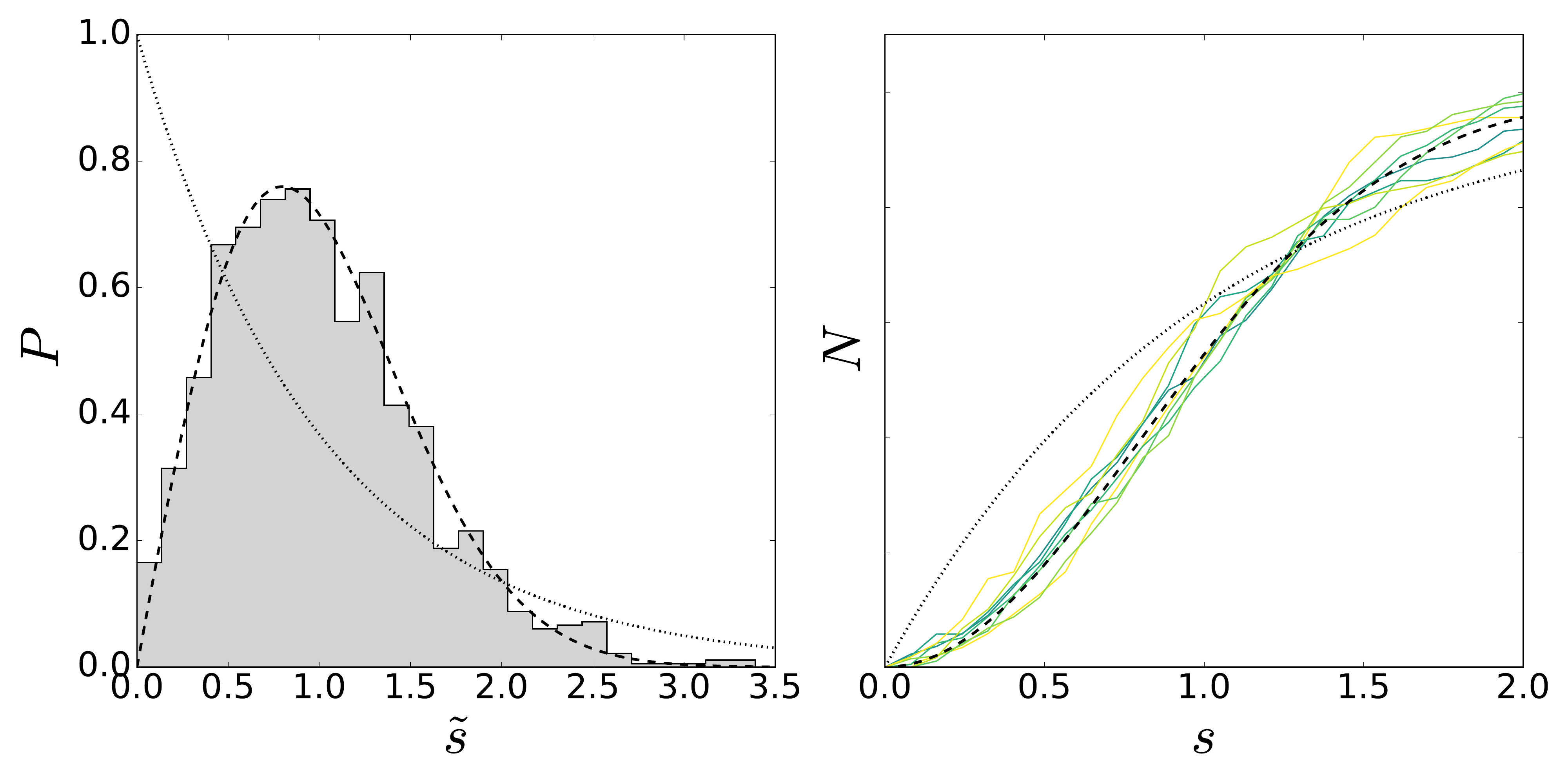}}
\caption{Left : nearest neighbors spacing distribution $P(\tilde{s})$,  with 1344 frequency levels obtained from eight independent spectra : $\Omega / \Omega_k = 0.481$ (206 odd levels), $\Omega / \Omega_k = 0.545$ (223 odd levels, 105 even levels), $\Omega / \Omega_k = 0.589$ (217 odd levels, 96 even levels), $\Omega / \Omega_k = 0.658$ (207 odd levels, 120 even levels) and  $\Omega / \Omega_k = 0.809$ (170 odd levels). Right : integrated distribution $N(s)$ for all eight independent spectra. In both panels the dashed line the Wigner surmise and the dotted line is the prediction for Poissonian spectra.}
\label{spacings}
\end{figure}

Our dataset of chaotic frequencies is described in table~\ref{table:1}. All axisymmetric modes were identified using the methodology described above. For non axisymmetric modes the last step, i. e. the identification of interface modes through the amplitude of the mode or the Husimi distribution, was not performed. The modes listed in table~\ref{table:1} are separated into symmetry classes associated with the quantum number m and the parity with respect to the equator ($m^-$ odd, $m^+$ even). A distinctive property of chaotic mode spectra is to have a universal statistical behavior \citep{gutzwiller_chaos_1990}. In particular, their nearest neighbor distribution $P(\tilde{s})$, with $\tilde{s}_n = (\omega_{n+1}-\omega_n) / \langle \omega_{n+1}-\omega_n \rangle$, where $\langle \omega_{n+1}-\omega_n \rangle$ corresponds to the mean frequency spacing, is expected to follow the Wigner surmise of Random Matrix Theory $P(\tilde{s}) = \frac{\pi \tilde{s}}{2}\, e^{- \pi s^2/4}$ \citep{gutzwiller_chaos_1990}. This is checked in Fig.~\ref{spacings} and it confirms that our method for selecting chaotic modes works properly. We aggregated the eight independent spectra to obtain better statistics (left panel). Using this standard aggregating procedure enables us to obtain much better  statistics than in previous studies \citep{lignieres_wave_2008}. The integrated distribution $N(s) = \int_0^{s}P(\tilde{s}) \, \R{d}\tilde{s}$, displayed in the right panel, shows that the agreement is good for individual spectra as well. We did not take into account the specific case of $\Omega / \Omega_k = 0.706$ which corresponds to an anomalous statistics, intermediate between the Wigner and Poisson distributions, with no level repulsion \citep{evano_correlations_2019}. Such a distribution is characteristic of a superposition of independent spectra, and will be explained later (see Section 4.2).

\begin{table}
\caption{Characteristics of chaotic frequency spectra. We show the symmetry classes, which are denoted by $m^{\pm}$ (where $m$ is the azimuthal quantum number, $+$ is even parity and $-$ is odd parity), the rotation rate, the number of levels and the frequency domain.}
\label{table:1}
\centering
\begin{tabular}{c c c c}
\hline\hline
symmetry class & Rotation ($\Omega / \Omega_k$) & levels & frequencies ($\omega / \omega_p$) \\
\hline
    $0^-$ & $0.481$ & 206 & 28.35 - 46.89 \\
    $0^-$ & $0.545$ & 223 & 28.15 - 44.09 \\
    $0^-$ & $0.589$ & 217 & 26.02 - 40.29 \\
    $0^-$ & $0.658$ & 207 & 36.37 - 44.89 \\
    $0^-$ & $0.706$ & 283 & 23.57 - 36.22 \\
    $0^-$ & $0.809$ & 170 & 24.02 - 30.01 \\
    $0^+$ & $0.545$ & 105 & 38.01 - 44.06 \\
    $0^+$ & $0.589$ & 96 & 30.52 - 36.60 \\
    $0^+$ & $0.658$ & 120 & 36.40 - 41.25 \\
    $1^-$ & $0.589$ & 125 &  30.51 - 38.48 \\
    $4^-$ & $0.589$ & 93 &  30.53 - 38.51 \\
\hline
\end{tabular}
\end{table}

%\begin{figure}[t]
%\centerline{\includegraphics[width=1\columnwidth]{interface_modes.png}}
%\caption{Examples of "interface modes" presenting characteristics of both regular and chaotic modes. Left panel : whispering gallery-like mode defined by its internal caustic. Right panel : Island-like mode centered around the stable island central orbit (black line). Top : level curves representing the rescaled pressure amplitude, whith the lower threshold set to 10\% of the maximum amplitude. Bottom : Pressure amplitude. \textcolor{blue}{Top/top/examples/new\_poly/script\_interface\_modes.py}}
%\label{interface_modes}
%\end{figure}

\section{Properties of high-frequency chaotic modes}
\label{sec:chaotic_modes_and_their_high_frequency_spectrum}

\subsection{Spatial distribution of the mode amplitudes}

\begin{figure}[!htbp]
\centerline{\includegraphics[width=1\columnwidth]{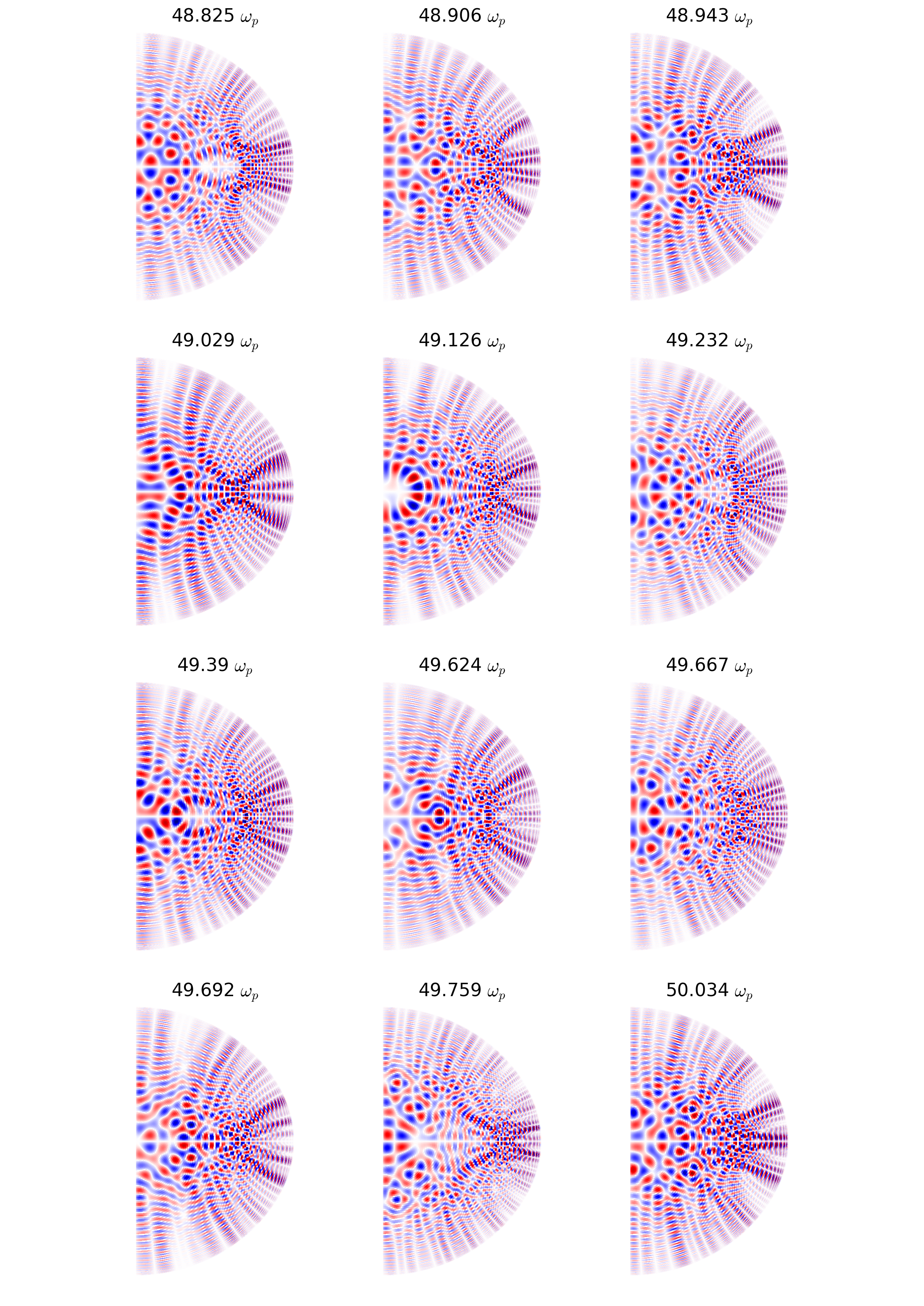}}
\caption{Scaled pressure amplitude $\Psi$ shown for twelve odd axisymmetric chaotic modes at $\Omega / \Omega_k = 0.589$ with quantum number $m=0$.}
\label{gallery_chaos}
\end{figure}

We start with a qualitative description of the main spatial features of chaotic modes. Representative examples of their amplitude distribution in a meridional plane are displayed in Fig.~\ref{gallery_chaos}. A first observation is that their nodal pattern is complex in the sense that, contrary to regular modes, there is no simple way to count the number of nodes. Nevertheless we also notice that while the inner part of the chaotic modes looks random, the outer part is much more structured. Indeed it is possible, near the surface, to count radial and angular nodes. The nodes appears regularly spaced in the radial direction but are unevenly distributed in the angular direction. The amplitude distribution of chaotic modes is therefore mainly characterized by its irregular nature, but with radial regularities near the surface. In the next subsection, we will show that the chaotic spectra present some regularities as well. Another notable property of the chaotic modes presented in Fig.~\ref{gallery_chaos} is that they spread out in the entire stellar interior. This distinguishes chaotic modes from regular modes that are confined in a narrow part of the star : whispering gallery modes stay close to the star surface and island modes are trapped in the vicinity of a central periodic orbit (see Fig.~\ref{4_modes}). Chaotic modes are thus the only class of p-modes able to probe the star center at high frequency.

% In those cases, the regular structure of the spectra is a reflection of the organized nature of the nodal patterns of the modes. In contrast, chaotic modes are characterized by their convoluted nodal pattern and their complex spectral structure with no apparent order.

\subsection{Regularities in the spectra}

\begin{figure}[!htbp]
\centerline{\includegraphics[width=1\columnwidth]{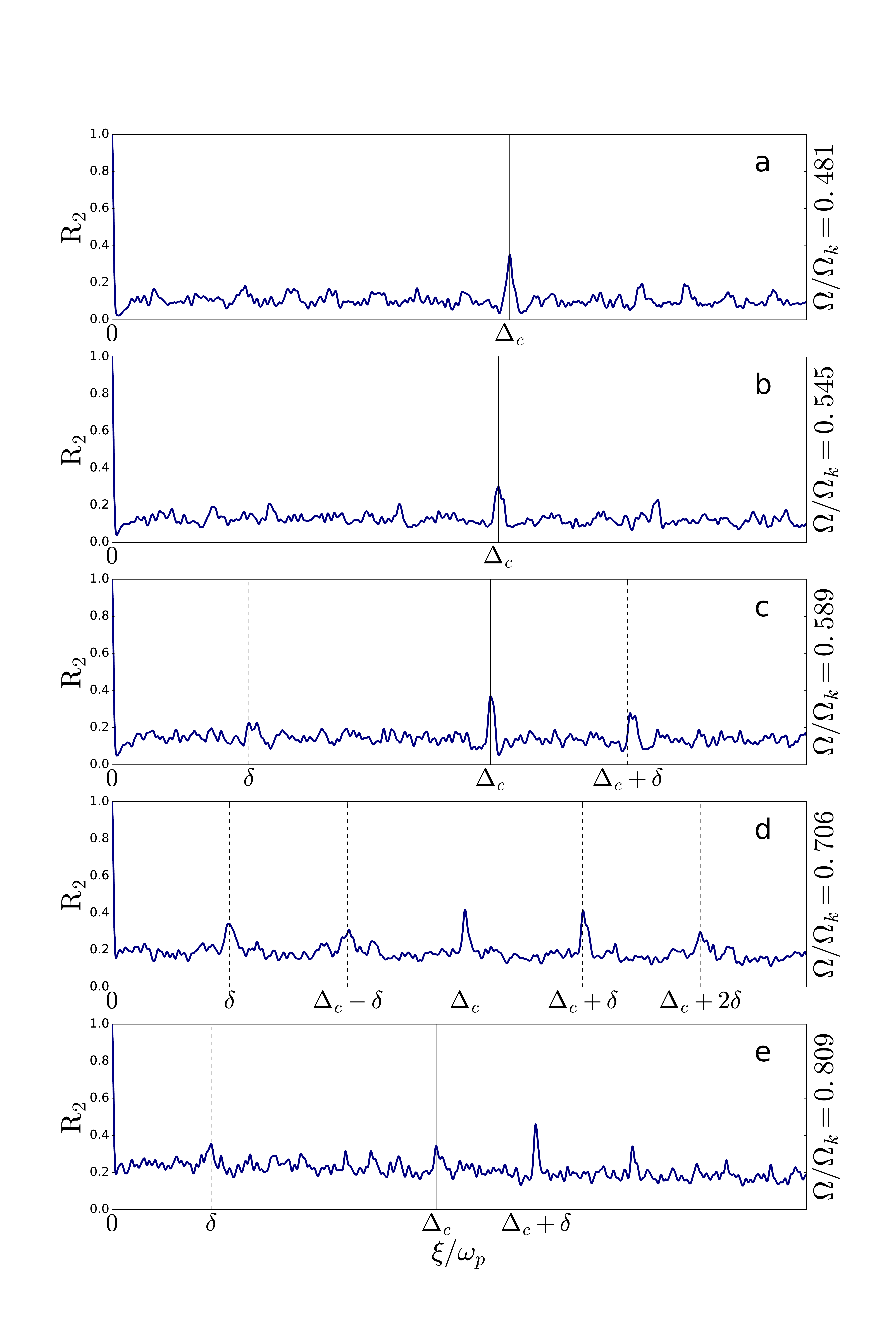}}
\caption{Autocorrelations $R_2(\xi)$, where $\xi$ is a displacement in frequency, of chaotic spectra with odd parity : a) 206 levels from 28.35 $\omega_p$ to 46.89 $\omega_p$, b) 223 levels from 28.15 $\omega_p$ to 44.09 $\omega_p$, c) 217 levels from 26.02 $\omega_p$ to 40.29 $\omega_p$, d) 283 levels from 23.57 $\omega_p$ to 36.22 $\omega_p$ and e) 170 levels from 24.02 $\omega_p$ to 30.01 $\omega_p$. The autocorrelations have been normalized such that their value at the origin is unity. The main peak's position is labeled $\Delta_c$ and marked with a solid line. Secondary peaks are marked in panels c, d and e with a dashed line.}
\label{autocorrelations}
\end{figure}

We computed the autocorrelation $R_2(\xi) = \int d(\omega - 1/2 \xi) d(\omega + 1/2 \xi) \, \R{d} \omega$ of a variety of chaotic spectra, where the density $d(\omega)$ was obtained by convoluting the spectra with a Gaussian function of small standard deviation compared to the mean frequency spacing and of height unity. The results are shown in Fig.~\ref{autocorrelations} at five different rotations for the case of axisymmetric modes with odd equatorial parity. The figure clearly shows peaks emerging from the noise level, which are not predicted by Random Matrix Theory nor seen in generic chaotic spectra. The peak that appears at every rotation, and which is usually the most visible, is referred to as the "main peak". Its position, denoted $\Delta_c$, slightly decreases with increasing rotation. Moreover, additional peaks of significant amplitude appear in the autocorrelations, their relative amplitude being large especially at $\Omega / \Omega_k = 0.706$. In the following we characterize the spectrum structure behind these peaks (the next two sub-sections) and then extend our analysis to the non-axisymmetric modes.

% Autocorrelation

\subsubsection{Main peak}
\label{sec:main_peak}

\begin{figure}[!htbp]
\centerline{\includegraphics[width=1\columnwidth]{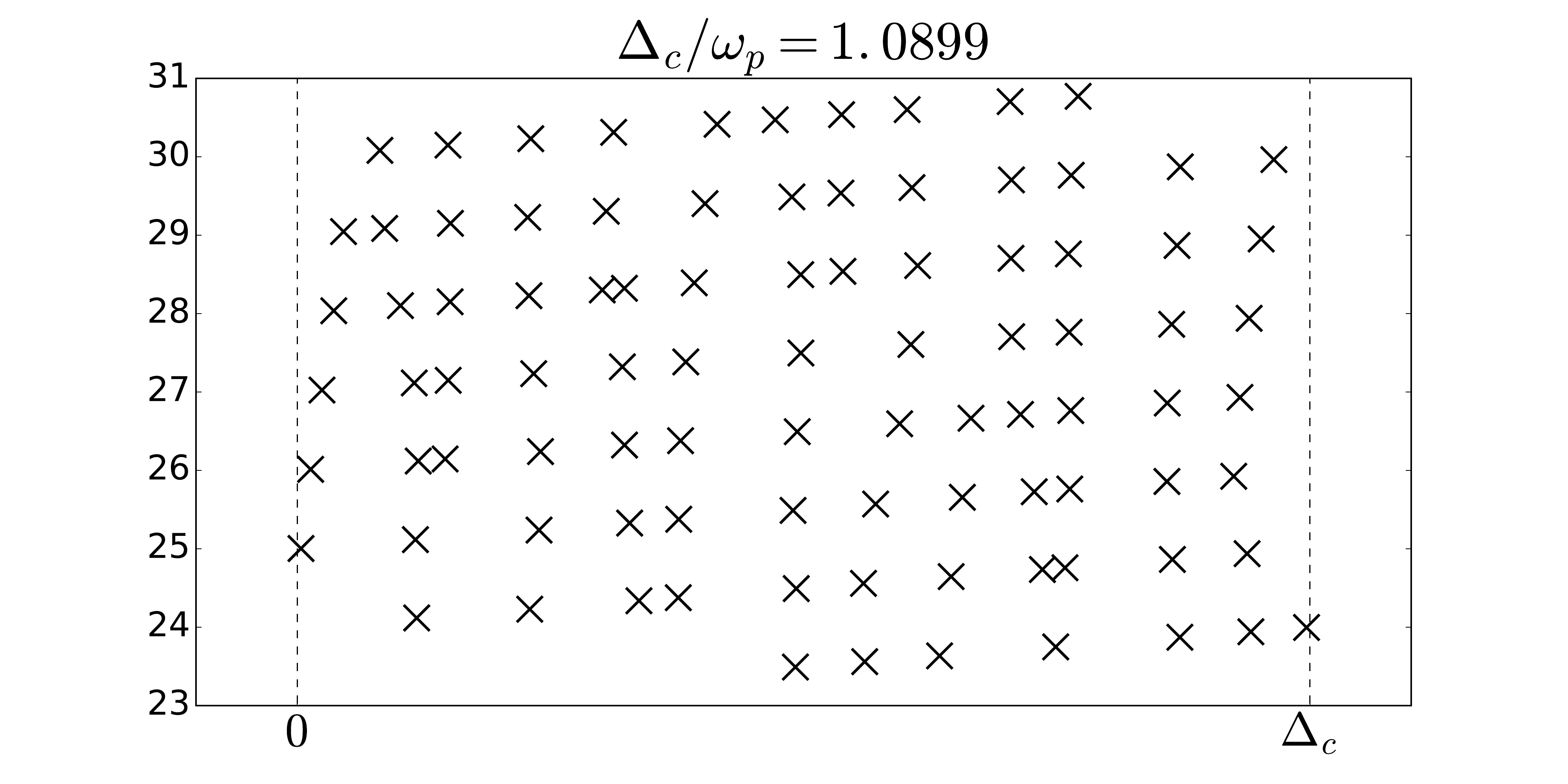}}
\caption{\'Echelle diagram of chaotic modes at $\Omega / \Omega_k = 0.589$ in the range $25.60\,\omega_p$ to $33.54 \, \omega_p$, with odd parity.}
\label{echelle_chaos}
\end{figure}

The systematic presence of the main peak in the autocorrelations is a hint that the spectra are structured by the characteristic frequency spacing $\Delta_c$. To verify this idea, we make use of so-called \'echelle diagrams. On such diagrams, two modes distant in frequency of $\Delta$ will be represented by two points sharing the same abscissa and separated by one unit on the y axis. We consider a portion of spectrum at $\Omega / \Omega_k = 0.589$, and analyze its structure in a detailed way. In Fig.~\ref{echelle_chaos}, the frequencies of all odd parity axisymmetric chaotic modes between $25.07 \, \omega_p$ and $33.78 \, \omega_p$ are represented on an \'echelle diagram, using a folding value equal to the main peak $\Delta_c = 1.0899 \, \omega_p$ of the autocorrelation. The frequencies are grouped in approximate vertical lines, some of which are very well aligned, whereas others seem to form diagonal or wavy tracks (Changing slightly the folding value breaks the best alignments but other tracks line up instead). The \'echelle diagram shows that the chaotic spectrum can be split into series of modes separated in frequency by approximately $\Delta_c$. \\

\begin{figure}[!htbp]
\centerline{\includegraphics[width=1\columnwidth]{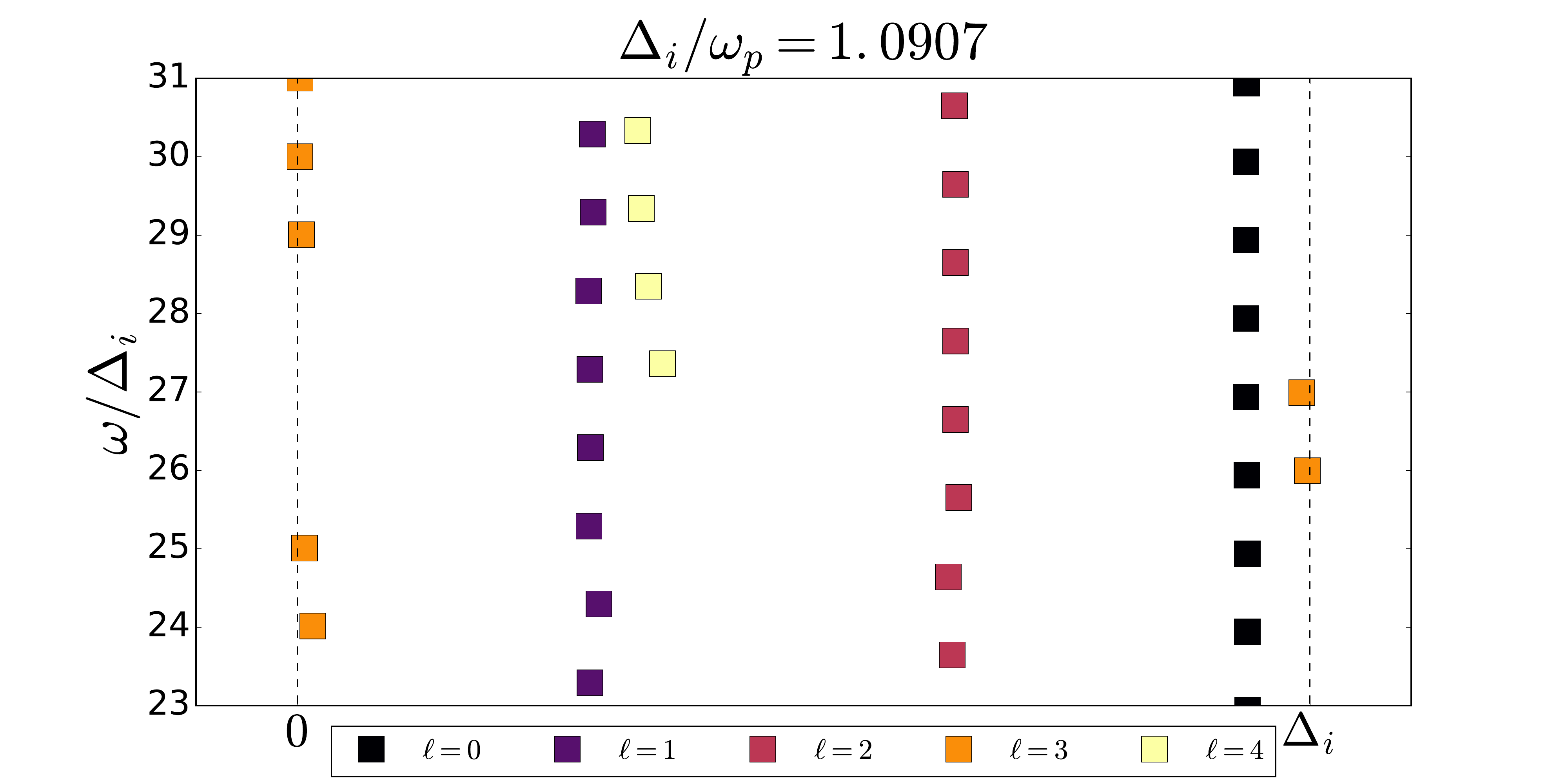}}
\caption{\'Echelle diagram of island modes $\ell = 0, 1, 2, 3, 4$ at $\Omega / \Omega_k = 0.589$, with only odd modes. The lowest point of the $\ell = 4$ track corresponds to an island-like interface mode.}
\label{echelle_islands}
\end{figure}

\noindent This property is well-known for low-degree modes in non-rotating stars as well as island modes in rotating ones (see Fig.~\ref{echelle_islands}). The frequency spacing between two consecutive modes is the so-called large separation and their amplitude distributions only differs by the number of nodes along a particular direction (radial for modes in a non rotating star and along the central periodic orbit for island modes). In such regular spectra, this structuring of the frequencies is directly related to the classical dynamics through so-called Einstein-Brillouin-Keller theory \citep{pasek_regular_2011,pasek_regular_2012}. Its appearance in a chaotic spectrum is more mysterious, and will be explained in Section 4.

In the autocorrelations of even chaotic mode spectra, both the main peak and secondary peaks usually have a smaller amplitude. Nonetheless, even mode frequencies produce ridges in the \'echelle diagram, just like odd mode frequencies. In \citet{lignieres_acoustic_2006}, it was pointed out that even modes are more strongly impacted by avoided crossings. This fact could explain why even modes are less regular than odd modes. \\

%\begin{figure}[t]
%\centerline{\includegraphics[width=1\columnwidth]{step-eps-converted-to.pdf}}
%\caption{Bottom : 217 chaotic frequencies at rotation $\Omega/\Omega_k = 0.589$. Top : spectral staircase function $\R{N}(\omega / \omega_p)$ couting the number of frequencies below $\omega / \omega_p$. The dashed line is a quadratic fit. \textcolor{red}{$%\R{N}(\omega)$ in spherical stars : Gough ?}}
%\label{step}
%\end{figure}

% Generically, the spatial distribution of chaotic modes is a Gaussian random speckle \textcolor{red}{(Ref. In the thesis of Claire Michel, Berry Houches 1981 is cited.)}. However, the chaotic modes of our dataset have two features that set them appart from the generic case. Firstly, we notice regions where the amplitude is enhanced, thus forming recognizable patterns. This phenomenon is reminiscent of the scars left by the smallest periodic orbits of the Hamiltonian system, but the patterns cannot be associated with a specific orbit in our case. Secondly, the nodal patterns appear more organized in the outter part of the star. In fact, close to the surface, the wavefronts are locally parallel to each other and it is possible to identify radial nodes.

\begin{figure}[!htbp]
\centerline{\includegraphics[width=1\columnwidth]{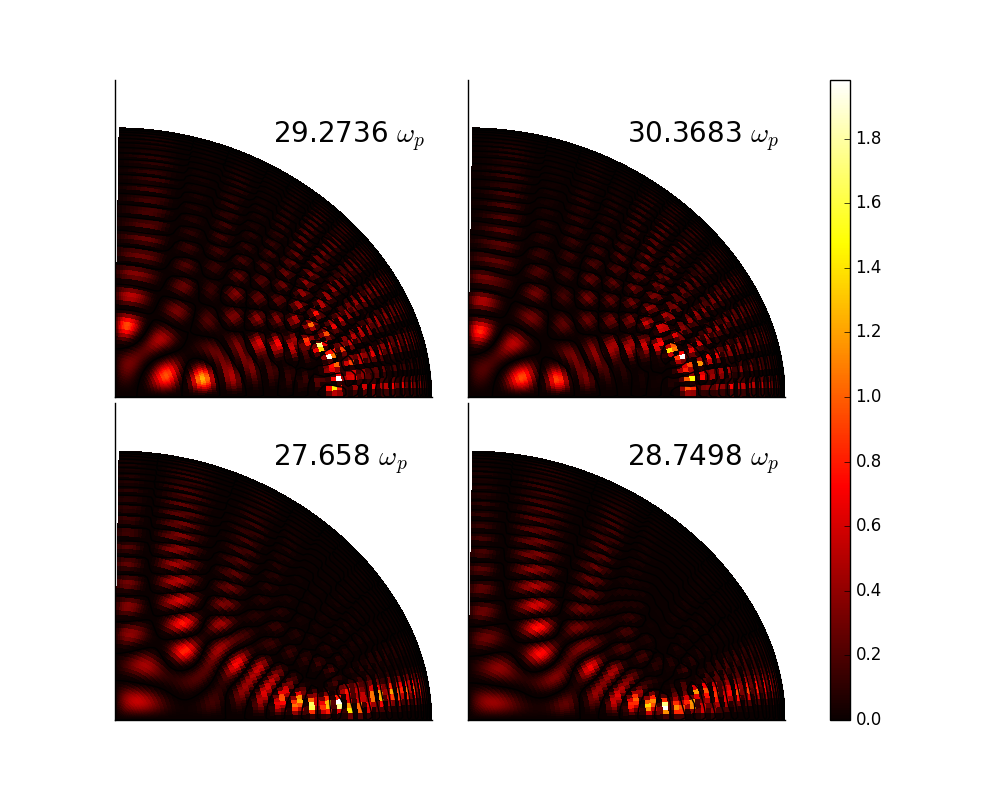}}
\caption{Mode intensity $|\Psi|^2$ at $\Omega/\Omega_k = 0.589$, where $\Psi$ is the scaled pressure amplitude, showing the similarity between consecutive modes. Top : two consecutive modes that belong to series 8. Bottom : two consecutive modes that belong to series 3.}
\label{consecutive}
\end{figure}

\begin{figure}[!htbp]
\centerline{\includegraphics[width=1\columnwidth]{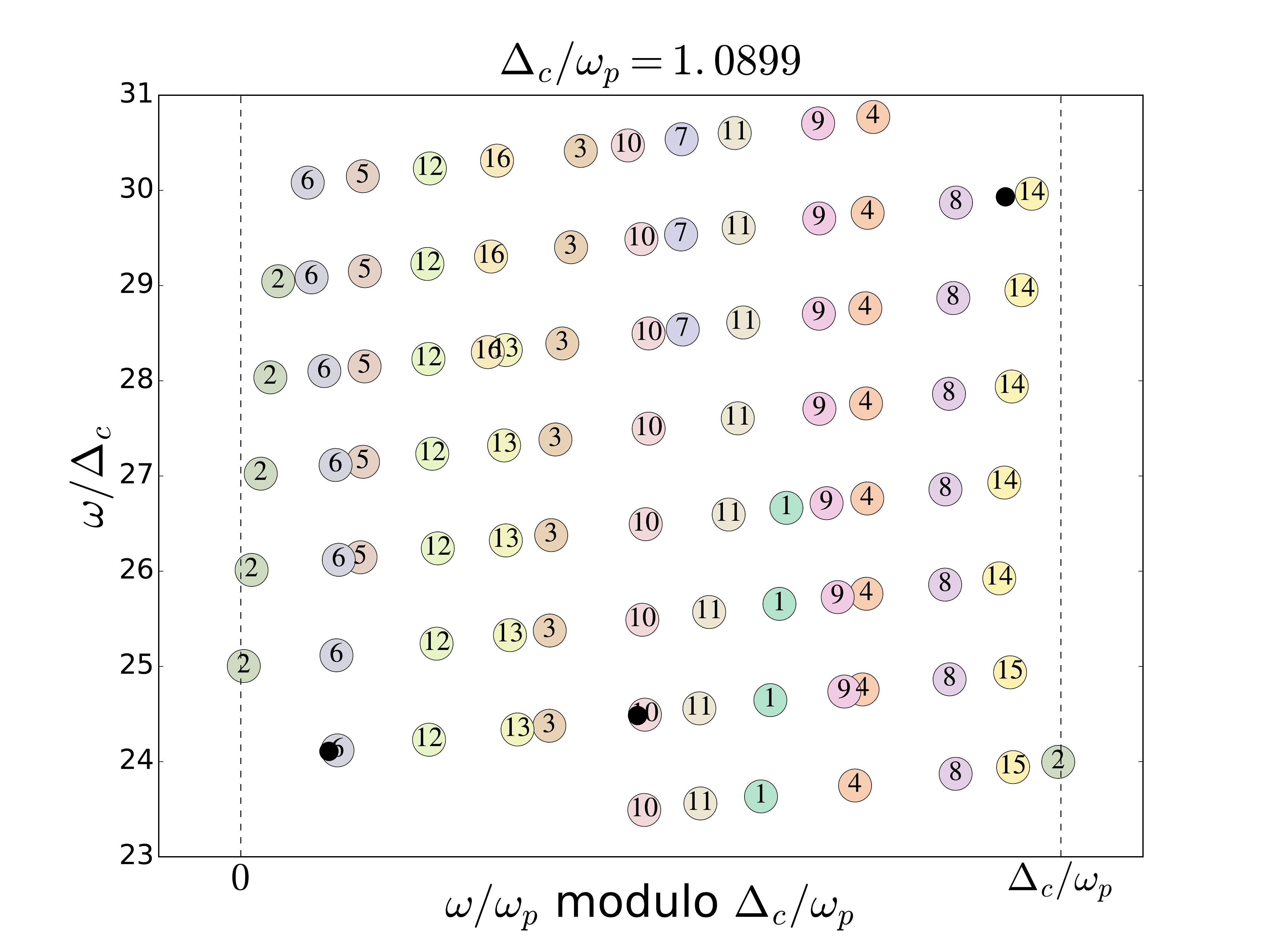}}
\caption{\'Echelle diagram showing all chaotic frequencies in the range $25.60\,\omega_p$ to $33.54 \, \omega_p$ modulo the large separation $\Delta_c$. Series of modes are labeled by numbers 1 to 16. Black dots correspond to modes that do not fit into a series.}
\label{echelle_2}
\end{figure}

 \noindent Comparing the amplitude patterns of the chaotic modes that belong to the same track on the \'echelle diagram, we find out that consecutive modes are often very similar. This is illustrated by two examples of consecutive modes in Fig~\ref{consecutive}. From a systematical search for couples of modes with a similar amplitude distribution that are separated in frequency by approximately $\Delta_c$ \footnote{the systematic search for consecutive modes was performed only for the chaotic spectrum at $\Omega/\Omega_k = 0.589$ represented in the \'echelle diagram.}, we were able to label the vertical series of chaotic mode \'echelle diagram, ending-up with the sixteen series shown in Fig.~\ref{echelle_2}. The very few modes left without a label are modes undergoing an avoided crossing.\\
 
\noindent This analysis indicates that the spectrum organization in series of modes separated by a fixed frequency spacing and showing similarities in their spatial distribution is also relevant for chaotic modes. In this context, $\Delta_c$ can be interpreted as a large separation for chaotic modes. Its value turns out to be close to the island mode large separation in the same frequency range $ \Delta_i = 1.0907 \omega_p$. \\

\noindent There are nevertheless some important differences with non-rotating and island modes. First the frequency spacing is much more regular for non-rotating or island modes than for chaotic modes. This is obvious from the comparison of the chaotic mode \'echelle diagram with the \'echelle diagram of the 2-period island modes shown in Fig.~\ref{echelle_islands}. The similarity of the amplitude distributions along a vertical series is also much stronger for island modes than for chaotic modes. In agreement with the asymptotic theory of island modes \citep{pasek_regular_2012}, island modes of the same series have the same number of nodes, denoted $\ell$, in the direction perpendicular to the periodic orbit. By contrast, the comparison between two chaotic modes of the same series but separated by a few $\Delta_c$ is not as clear, since the patterns slowly evolves from one mode to the next. Another important difference comes from the characteristics of the series. As expected from the theory and as observed in Fig.~\ref{echelle_islands}, island mode series should not stop toward high frequencies as modes with the same $\ell$ and higher $n$ remain aligned in the \'echelle diagram. This is not the case for chaotic modes as some series like series number 1, 2, and 13 on Fig.~\ref{echelle_2} come to an end in the frequency range considered. In parallel, some series starts above a given frequency, for example the series number 16 appearing above $\omega = 27 \Delta_c$.  Typically, interface modes appear at the start of a series as whispering gallery-like modes or at the end of a series as island-like modes. In our data, the end of a series of chaotic modes is also the start of a series of island modes of given $\ell$ ; for instance, series 1 is followed by $\ell = 4$ island modes and series 2 by $\ell = 5$ island modes. This transition is visible by comparing Fig.~\ref{echelle_2}, where series 1 ends just below $27 \Delta_c$, and Fig.~\ref{echelle_islands} where the $\ell = 4$ island modes start just above $27 \Delta_c$. \\ 

\subsubsection{Secondary peaks}
\label{sec:secondary_peaks}

In addition to the main peak discussed above, we see other peaks that we call secondary peaks. At $\Omega/\Omega_k = 0.589$, $\Omega/\Omega_k = 0.706$ and $\Omega/\Omega_k = 0.809$, one can clearly distinguish these peaks from the background noise. They are marked by a dashed line in Fig.~\ref{autocorrelations}. The peaks occur at $\alpha \delta + \beta \Delta_c$, where $\delta$ is the position of the first secondary peak and $\alpha$ and $\beta$ are integers. For instance, the peaks of Fig.~\ref{autocorrelations}, panel c, occur at $(\alpha = +1, \, \beta = 0)$ and $(\alpha = +1, \, \beta = +1)$. At $\Omega/\Omega_k = 0.706$, secondary peaks are numerous and close in amplitude to the main peak. At this rotation, we have already noticed the nearest neighbors statistics is anomalous \citep{evano_correlations_2019}, intermediate between the Wigner and Poisson distributions with no level repulsion. This anomalous statistics is a sign that independent subspectra coexist at $\Omega / \Omega_k = 0.706$. We will propose an explanation for these secondary peaks in Section 4.

%The échelle diagram for odd modes at $\Omega/\Omega_k = 0.706$, with a $\delta$ folding, is shown in Fig.~\ref{echelle_71}. As in the previous section, we intended to identify families of modes using two criteria : the modes are regularly spaced in frequency and consecutive modes have a similar amplitude distribution. Following this methodology, we found two families of modes highlighted on the échelle diagram while no clear pattern similarities have been detected among the other modes.

%\begin{figure}[t]
%\centerline{\includegraphics[width=1\columnwidth]{echelle_71-eps-converted-to.pdf}}
%\caption{Échelle diagram at $\Omega / \Omega_k = 0.706$ with a folding value of $\Delta = 1.0505 \omega_p$. \textcolor{blue}%{echelle\_figure.py}}
%\label{echelle_71}
%\end{figure}

%\begin{figure}[t]
%\centerline{\includegraphics[width=1\columnwidth]{echelle_71-eps-converted-to.pdf}}
%\caption{Échelle diagram at $\Omega / \Omega_k = 0.706$ with a folding value of $\delta = 0.3385 \omega_p$. \textcolor{blue}{echelle\_figure.py}}
%\label{echelle_71}
%\end{figure}

\subsection{Non-axisymmetric modes}

\begin{figure}[!htbp]
\centerline{\includegraphics[width=1\columnwidth]{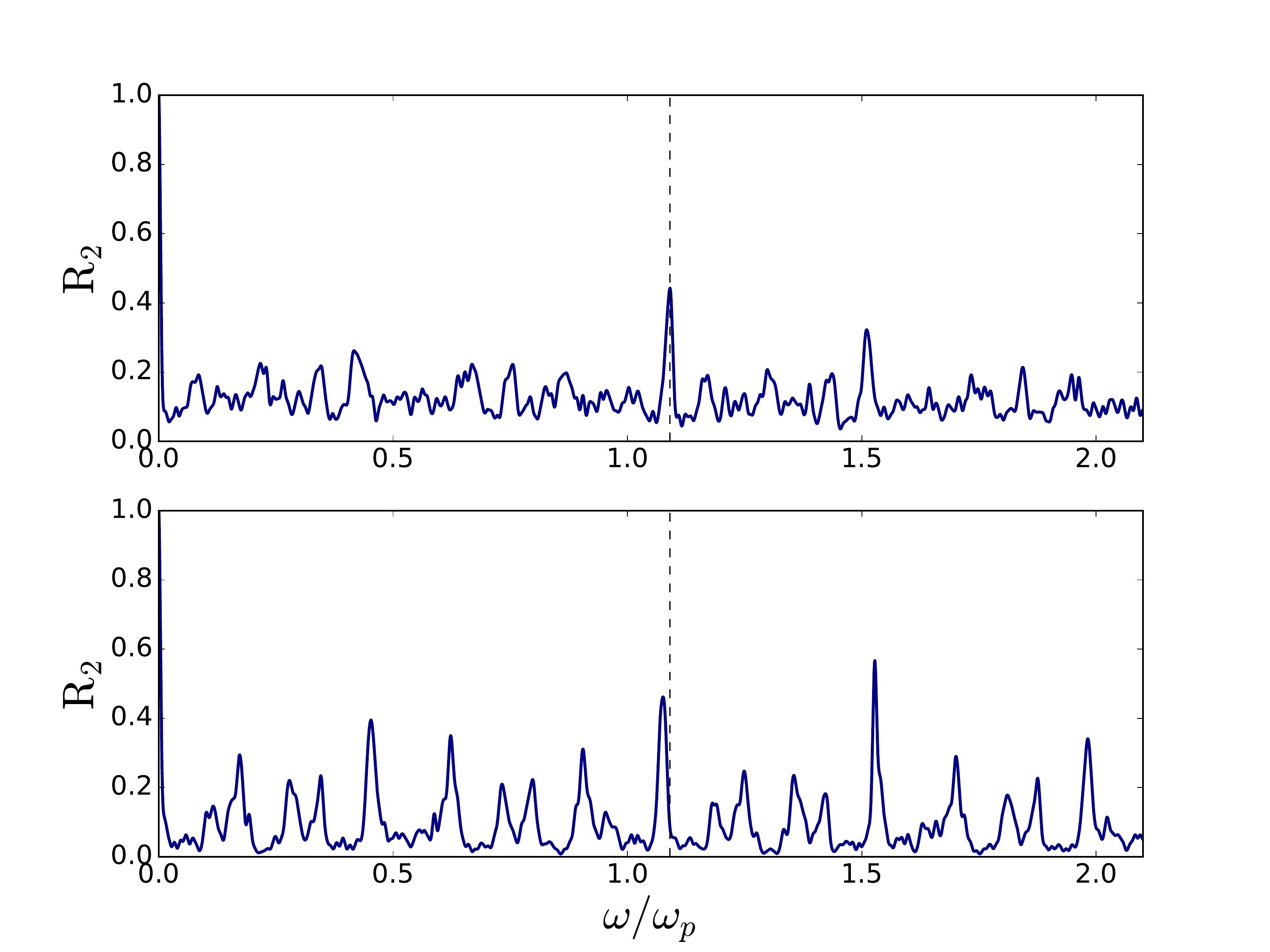}}
\caption{Top panel : Autocorrelation at $\Omega / \omega_k = 0.589$ for quantum number $m = 1$, in the frequency domain $30.51 \omega_p$ to $38.48 \omega_p$. Bottom panel : Autocorrelation at $\Omega / \omega_k = 0.589$ for quantum number $m = 4$, in the frequency domain $30.53 \omega_p$ to $38.51 \omega_p$. In both cases, the dashed line is the position $\Delta_c$ of the main peak for axisymmetric modes at the same rotation rate.}
\label{auto_m4}
\end{figure}

Because the model is cylindrically symmetric, modes are quantized in the azimuthal direction, the quantum number being denoted $m$. In the ray dynamics, it corresponds to the quantization of the invariant associated with this symmetry, that is the projection $L_z$ of the angular momentum on the axis of rotation. For a given rotation and frequency, increasing $\tilde{L}_z = L_z/\omega$ has the effect of reducing the size of the chaotic zone. This can be seen by comparing the PSS of Fig.~\ref{PSS_59} with $m = 0$ and the PSS of Fig.~\ref{PSS_59_m4} with $m=4$, both with $\omega = 24.41 \omega_p$. From the mode numerical computations, we find for $m = 1$ the main peak is still clearly visible in the autocorrelation, and almost exactly at the same position as for the axisymmetric case. However for $m=4$ the autocorrelation shows a forest of peaks, the main peak being slightly shifted toward low values (see Fig.~\ref{auto_m4}).

\section{Semiclassical interpretation}
\label{sec:semiclassical_analysis}

\begin{figure}[!htbp]
\centerline{\includegraphics[width=1\columnwidth]{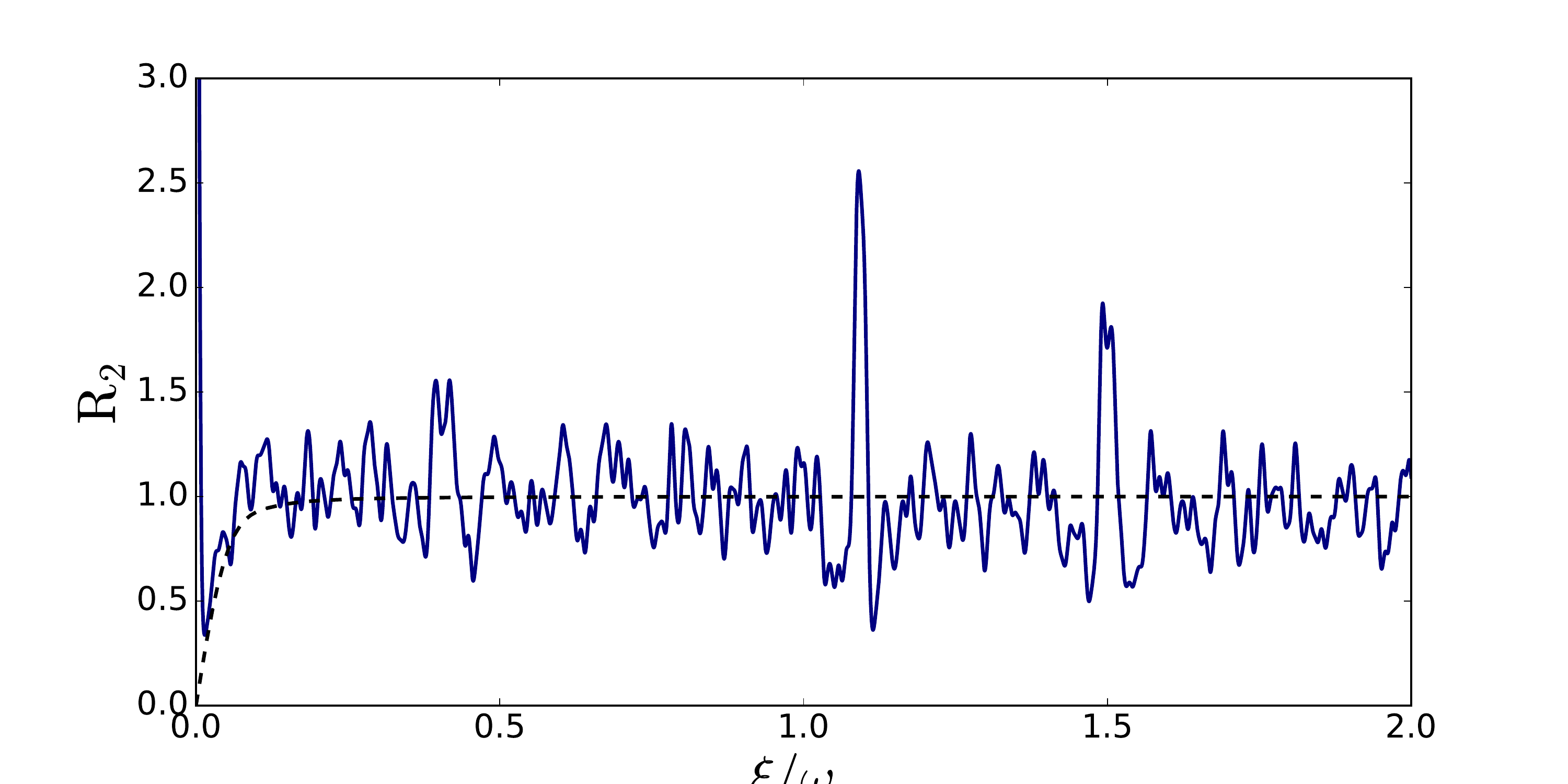}}
\caption{Comparison of the "stellar" autocorrelation $R_2(\xi)$ at $\Omega / \Omega_k = 0.589$, represented in solid line, with the GOE autocorrelation represented by a dashed line. To compare the "stellar" and GOE autocorrelations, two changes have been made. First, the autocorrelation of our model has been re-scaled so that the value of the plateau, i.e. the line around which there are fluctuations, is approximately unity. The value of the plateau otherwise depends on the mean density $\bar{d}(\omega) = \R{d} \bar{N} / \R{d} \omega$ and the Gaussian smoothing of the spectrum. Secondly, the GOE distribution has been rescaled, in the direction of the x axis, by the mean level spacing of the $\Omega / \Omega_k = 0.589$ spectrum in the frequency domain considered.}
\label{auto_GOE}
\end{figure}

In section \ref{sec:chaotic_modes_and_their_high_frequency_spectrum}, we gave a description of the most important properties of chaotic modes in our model of rotating stars. Our goal in the present section is to provide a theoretical understanding of these results based on asymptotic methods. The statistical properties of chaotic spectra are well described by Random Matrix Theory \citep{bohigas_houches_1989, metha_random_2004}, which studies the distribution of the eigenvalues of matrices filled with Gaussianly distributed random numbers. Real symmetric random matrices are of particular interest. Indeed, they form the Gaussian Orthogonal Ensemble (GOE) that models the spectra of time-reversal symmetric systems. Comparing an autocorrelation of the stellar model with the GOE autocorrelation %\footnote{The GOE autocorrelation is expressed as $R_{2, \, GOE}(\xi) = \bar{d}^2(1-Y_2(\xi))$. Where $\bar{d}$ is a locally averaged frequency density and $Y_2(\xi)$ is the two-points cluster function \citep{bohigas_houches_1989}} 
(see Fig.~\ref{auto_GOE}) makes it clear that the main peak and secondary peaks are not generic features of wave chaos. If Random Matrix Theory is efficient at modeling the generic properties of chaotic systems, it is not able to grasp specific behaviors that may arise in particular systems (see e.g., \citet{arithm_1992,arithm_1997}). However, semiclassical methods based on the propagation of rays are well suited for this task. \\

Using the semiclassical periodic orbit theory, which relates the mode properties to the acoustic ray dynamics, we show in subsection \ref{main_peak} that the $\Delta_c$ regularity is caused by the strong decrease of the sound speed at the surface, and present a theory which predicts the occurrence of the peak and its characteristics from the ray dynamics. We then explain the presence of secondary peaks by the transport properties of the phase space in subsection \ref{secondary_peaks_2}. In subsection \ref{amplitude} we show, using a simplified model, how the behavior of acoustic rays near the surface may induce structure in the nodal pattern. In subsection \ref{families} we discuss how the families of chaotic modes evolve when the frequency domain changes. At last in subsection \ref{sec:chaos_vs_islands}, we show that the proximity of $\Delta_c$ and $\Delta_i$ is not accidental, and we propose a way to differentiate chaotic modes and island modes using the symmetries of the system. \\

\subsection{Main autocorrelation peak}
\label{main_peak}

\subsubsection{Periodic orbit theory}

Before diving into the subject of periodic orbit theory, let us introduce a few quantities that will be later used to characterize the statistical properties of the spectra. The staircase function $N(\omega)$ counts the number of modes below a given frequency $\omega$. From the staircase function one can define the spectral density $\R{d}(\omega) = \R{d}N/\R{d} \omega$. Both the staircase function and the spectral density are often written as the sum of two contributions called the smooth part $\bar{N}(\omega)$ (or $\bar{\R{d}}(\omega)$) and the fluctuating (or oscillating) part $N^{osc}(\omega)$ (or $\R{d}^{osc}(\omega)$), where the smooth part is obtained by locally averaging the function in the neighborhood of a target frequency.\\

\noindent Periodic orbit theory is an asymptotic semiclassical theory developed in the limit of high frequency or short wavelength. It is based on the trace formula \citep{gutzwiller_chaos_1990}, which relates the spectral density to a sum over the periodic trajectories of the Hamiltonian system. Though it was originally derived in the framework of quantum mechanics, the trace formula can be adapted to any wave system with a ray approximation. Indeed trace formulas have been built and tested, for instance in optics using microwave cavities \citep{kudrolli_signatures_1994} or for plate vibrations \citep{Hugues_98}. In the same spirit, we re-derive the trace formula for the system considered here in Appendix \ref{trace_formula}. Using $j$ as a label for the periodic orbits $\gamma_j$, the formula reads :

\begin{equation}
    \label{gutzwiller}
    \R{d}(\omega)-\R{\bar{d}}(\omega) = \mathrm{Re}\sum_j A_j \, e^{\R{i} S_j (\omega)},
\end{equation}

\noindent where $S_j = \omega \oint_{\gamma_j} \R{d}s / \tilde{c}_s$ is the action and $\oint_{\gamma_j} \R{d}s / \tilde{c}_s = T_j$ is the acoustic travel time of $\gamma_j$. The amplitude term is expressed as

\begin{equation}
    A_j = \frac{1}{\pi} \frac{T_j}{|\det (M_j-I)|^{1/2}},
\end{equation}

\noindent where $I$ denotes the identity matrix and $M_j$ is the monodromy matrix which describes the linearized motion around the periodic orbit and whose eigenvalues give the stability of the orbit. The periodic orbits are either so-called primitive orbits or repetitions of them. In many systems, the density $\rho(T)$ of orbits with a time period $T$ grows exponentially as

\begin{equation}
\label{eq:density}
\rho(T) \approx (1/T) \, e^{\lambda T},
\end{equation}

\noindent where $\lambda$ is the average lyapunov exponent of the system, which describes the rate at which nearby trajectories diverge. Long orbits are less stable and their amplitude drops down as

\begin{equation}
\label{eq:amplitude}
A(T) \approx \frac{1}{\pi} \, T \, e^{-(1/2)\lambda T}.
\end{equation}

The exponential growth of the number of long orbits makes it numerically difficult to attain good precision on the spectral density using the trace formula. Moreover in many systems, including the stellar model studied here, there is no automatic procedure to find the periodic orbits. To circumvent these issues, Berry showed \citep{berry_semiclassical_1985} that some statistical quantities such as the variance, or more importantly the autocorrelation of the spectrum, can be approximated using the trace formula in systems where individual periodic orbits are not known, provided one knows their distribution. \\

The autocorrelation $R_2(\xi) = \left \langle \R{d}(\omega - 1/2 \, \xi)\, \R{d}(\omega + 1/2 \, \xi) \right \rangle$, with the average $\langle f \rangle = \int f \R{d} \omega$, can be re-written as $R_2(\xi) = \left<\R{d^{osc}}(\omega-1/2 \xi) \, \R{d^{osc}}(\omega+1/2 \xi) \right\rangle + \phi(\xi)$, where $\phi(\xi)$ is a smooth function. The theory aims at deriving an approximate expression of the form factor :

\begin{equation}
    K(T) = \frac{1}{\sqrt{2 \pi}}\int^{\infty}_{-\infty} \R{d} \xi \, \R{exp}(\R{i} \xi T) \, \R{C}(\xi),
    \label{form_factor}
\end{equation}

\noindent which is the Fourier transform of the autocorrelation

\begin{align}
C(\xi) &= \left<\R{d^{osc}}(\omega-1/2 \xi) \, \R{d^{osc}}(\omega+1/2 \xi) \right\rangle, \label{eq:c_of_xi} \\
       &= \Big \langle \Big(\R{Re} \sum_i A_i e^{\R{i} S_i(\omega - 1/2 \xi)}\Big) \Big(\R{Re} \sum_j A_j e^{\R{i} S_j(\omega + 1/2 \xi)}\Big) \Big \rangle.
\end{align}

\noindent Getting rid of the real parts through the fact that with $x = \R{Re}(\tilde{x}) = A_i \cos(\xi T_i + S_i)$ and $y = \R{Re}(\tilde{y}) = A_j \cos(\xi T_j + S_j)$ then $\langle x \, y \rangle = 1/2 \, \langle \tilde{x} \, \tilde{y}^* \rangle$, gives

\begin{equation}
C(\xi) = \frac{1}{2}\Big \langle \Big(\sum_i A_i e^{\R{i} S_i(\omega - 1/2 \xi)}\Big) \Big(\sum_j A_j e^{\R{i} S_j(\omega + 1/2 \xi)}\Big)^* \Big \rangle.
\end{equation}

\noindent Inserting this expression in Eq.~\ref{form_factor} and expanding the action as $S_j(\omega \pm \, 1/2 \xi) \approx S_j(\omega) \pm 1/2 \, \xi \, (\partial S_j / \partial \omega) =  S_j(\omega) \pm 1/2 \, \xi T_j(\omega)$, where $T_j$ is the acoustic travel time introduced before, leads to a new expression of the form factor :

\begin{equation}
\begin{aligned}
K(T) &= \frac{1}{2} \left \langle \sum_i \sum_j A_i A_j^* \, e^{i(S_i-S_j)} \, \frac{1}{\sqrt{2 \pi}} \int_{-\infty}^\infty \R{d}\xi \, e^{i \xi  (T-\frac{1}{2}(T_i+T_j))} \right \rangle, \\
    &= \frac{1}{2} \left \langle \sum_i \sum_j A_i A_j^* \, e^{i(S_i-S_j)} \, \frac{2 \pi}{\sqrt{2 \pi}}\delta \Big(T - \frac{1}{2}(T_i+T_j)\Big) \right \rangle.
\end{aligned}
\end{equation}

\noindent Under the frequency average $\langle.\rangle$, the contribution of the off-diagonal terms $i \neq j$ in the double sum can be neglected. This diagonal approximation is valid for "short times" below the Ehrenfest time $T_E \approx (1/\lambda) \ln(\omega)$ \citep{arithm_1997}. Beyond $T_E$, there are pairs of orbits with very close actions $S_i \approx S_j$ and higher order terms need to be computed \citep{bogomolny_gutzwillers_1996, sieber_correlations_2001}. In the regime where the diagonal approximation is valid, the expression of the form factor is reduced to

\begin{equation}
    \label{eq:k_T}
    \R{K}(T) \approx \sum_j A_j^2 \delta(T-T_j).
\end{equation}

\noindent In this expression $K(T)$, as a Fourier transform of a correlation function in frequency, is a function of time $T$. Eq.~\ref{eq:k_T} means that $K(T)$ is related to the distribution of travel times of all the periodic orbits labeled by $j$ (and not of their lengths). This will be crucial in the subsequent analysis. For long orbits Eq.~\ref{eq:k_T} can be written as 

\begin{equation}
K(T) \approx A^2(T) \, \rho(T),
\end{equation}

\noindent where $\rho(T) \, \R{d}T$ is the the number of periodic orbit with a time period between $T$ and $T+\R{d}T$. A prescription for the amplitude $A(T)$ and the density $\rho(T)$ is given in Eq.~\ref{eq:amplitude} and Eq.~\ref{eq:density}, and can be generalized \citep{Hannay_84}, leading to the conclusion that for generic chaotic systems \citep{berry_semiclassical_1985} the form factor is a linear function of $T$ i.e.  $K(T) \propto T$ (up to the Ehrenfest time), which is in accordance with the predictions of GOE \citep{bohigas_houches_1989}. Such a result does not predict the occurrence of peaks in the autocorrelation. We will show that the same theory but using the specific  density of periodic orbits of our system leads to a different result predicting autocorrelation peaks.

\subsubsection{Distribution of acoustic travel times}

To this aim, we need to model the density $\rho(T)$ of periodic orbit travel times in our system. First, we define a chord as a portion of orbit between two consecutive rebounds at the surface. From this definition, a trajectory that bounces $n$ times at the surface will be called a $n-chord$. For any integer $n$, periodic trajectories with n rebounds are a subset of n-chord trajectories. As already mentioned, we do not know any systematic way of finding all the periodic orbits in the system. Nonetheless, we may infer some of their properties by studying large samples of n-chord trajectories.\\

\begin{figure}[!htbp]
\centerline{\includegraphics[width=1\columnwidth]{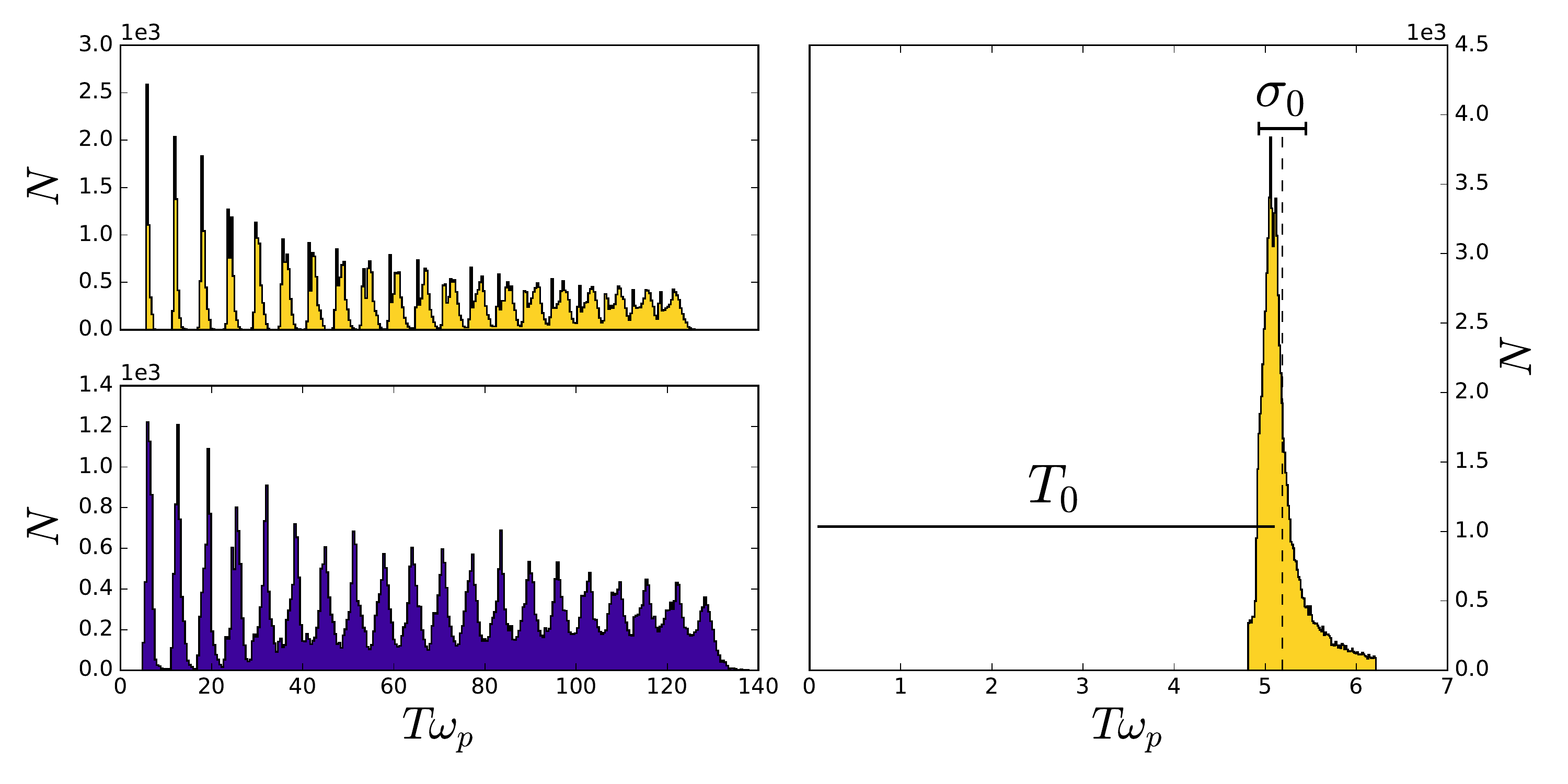}}
\caption{Left panels : Number of n-chord trajectories, with n = 1, ..., 20, vs their travel time $T$ at rotations $\Omega / \Omega_k = 0.481$ (top) and $\Omega / \Omega_k = 0.809$ (bottom), with 300 bins for the total distribution. The n-chord samples contain $\sim 4200$ chords each. Right panel : the $\Omega / \Omega_k = 0.481$ 1-chord distribution in more details, with $\sim 84000$ chords and $100$ bins. The mean value $T_0$ of the distribution is marked with a dashed line and the standard deviation $\sigma_0$ is shown.}
\label{distribution}
\end{figure}

At a given rotation rate, we compute the acoustic travel times of thousands of n-chord trajectories. The 1-chord distribution is shown in the right panel of Fig.~\ref{distribution}, at $\Omega / \Omega_k = 0.481$. The distribution is a narrow packet : its standard deviation $\sigma_0$ is small compared to its mean value $T_0$. We interpret this effect as a consequence of the strong decrease of the sound speed near the surface \footnote{To test this hypothesis, we computed the acoustic time distribution in a domain shaped like a $\Omega / \Omega_k = 0.706$ rotating star, but with a homogeneous sound speed throughout the interior. It results in a distribution whose dispersion is of the same order as the mean value : $\sigma_0 \approx 0.32 \, T_0$.}. Indeed, the trajectories travel rapidly through the core of the star and the acoustic time is dominated by the surface behavior, so the actual length of the trajectory has very little impact on its travel time. We expect the characteristic time $T_0$ to increase with rotation, because of the increasing volume of the star. This is indeed confirmed, with $T_0 = 5.19 / \omega_p$ at $\Omega/\Omega_k = 0.481$ and $T_0 = 6.30 / \omega_p$ at $\Omega/\Omega_k = 0.706$. \\

\noindent Moreover, due to the centrifugal deformation, the region where the sound speed is very small widens at the equator. Hence the dispersion of travel times must increase as well with rotation, as can be seen by comparing, for instance, the chord joining the two poles with the chord joining opposite points on the equator. The total distribution of n-chords with n $\le 20$, with the same number of orbits for all values of n, is represented in Fig.~\ref{distribution}, panels (a) and (b) at rotations $\Omega / \Omega_k = 0.481$ and $\Omega / \Omega_k = 0.809$. The packets are evenly spaced out ($T_n \approx n T_0$) but get thicker as n increases ($\sigma_n \approx \sqrt{n} \sigma_0$), since each packet can be seen as the sum of $n$ independent variables. Thus this packet structure will disappear when $n$ becomes large, i.e. for long travel times. An important difference between the two distributions shown in Fig.~\ref{distribution} is the rate at which adjacent packets overlap, leading to the disappearance of the packet structure. This is quantified by the ratio $\sigma_0 / T_0$ which is $0.049$ at $\Omega / \Omega_k = 0.481$ and $0.097$ at $\Omega / \Omega_k = 0.809$. \\

\noindent The constraints imposed by the n-chord acoustic time distribution could be strong enough to impose a kind of periodic oscillation in the distribution of periodic orbits, and thus in the form factor $K(T)$. This oscillation could then produce a peak in the autocorrelation in virtue of the Fourier relation \footnote{Looking closely, one can discern a very narrow peak inside each packet. It is created by trajectories trapped near the main island (see subsection \ref{secondary_peaks_2}) and will have no significant impact on the autocorrelation}. We will now show that it is indeed the case \\

To this aim, we model the distribution of travel times as a sum of Gaussian functions : $\R{P_{\, \Omega}}(T) = \sum_n \R{P_{n, \, \Omega}}(T)$, with $\R{P_{n, \, \Omega}}(T)$ the probability distribution of n-chords travel times :

\begin{equation}
    \R{P_{n, \, \Omega}}(T) = \frac{T_0}{\sqrt{2 \pi n} \, \sigma_0} \, \R{exp} \left( - \frac{(T-n \, T_0)^2}{2 (\sqrt{n} \, \sigma_0)^2} \right),
    \label{podist}
\end{equation}

\noindent where $T_0$, in the prefactor, has been added for normalization purposes. It is a crude approximation but it encapsulates in a simple form all the relevant properties of the distributions, namely the presence of regularly spaced packets and their widening with increasing $n$. One has to keep in mind that the dependency on $\Omega$ is not explicit but hidden in the values of $T_0$ and $\sigma_0$. Finally, we need to take into account the fact that the number of trajectories of acoustic time $T$ grows with $T$. Thus, we write $\rho(T) = (1/T) e^{\lambda T} \times \R{P_{\, \Omega}}(T)$ and the form factor becomes $K(T) \propto A^2(T)\; \rho(T) = T \, \R{P_{\, \Omega}}(T)$. \\

\begin{figure}[!htbp]
\centerline{\includegraphics[width=1\columnwidth]{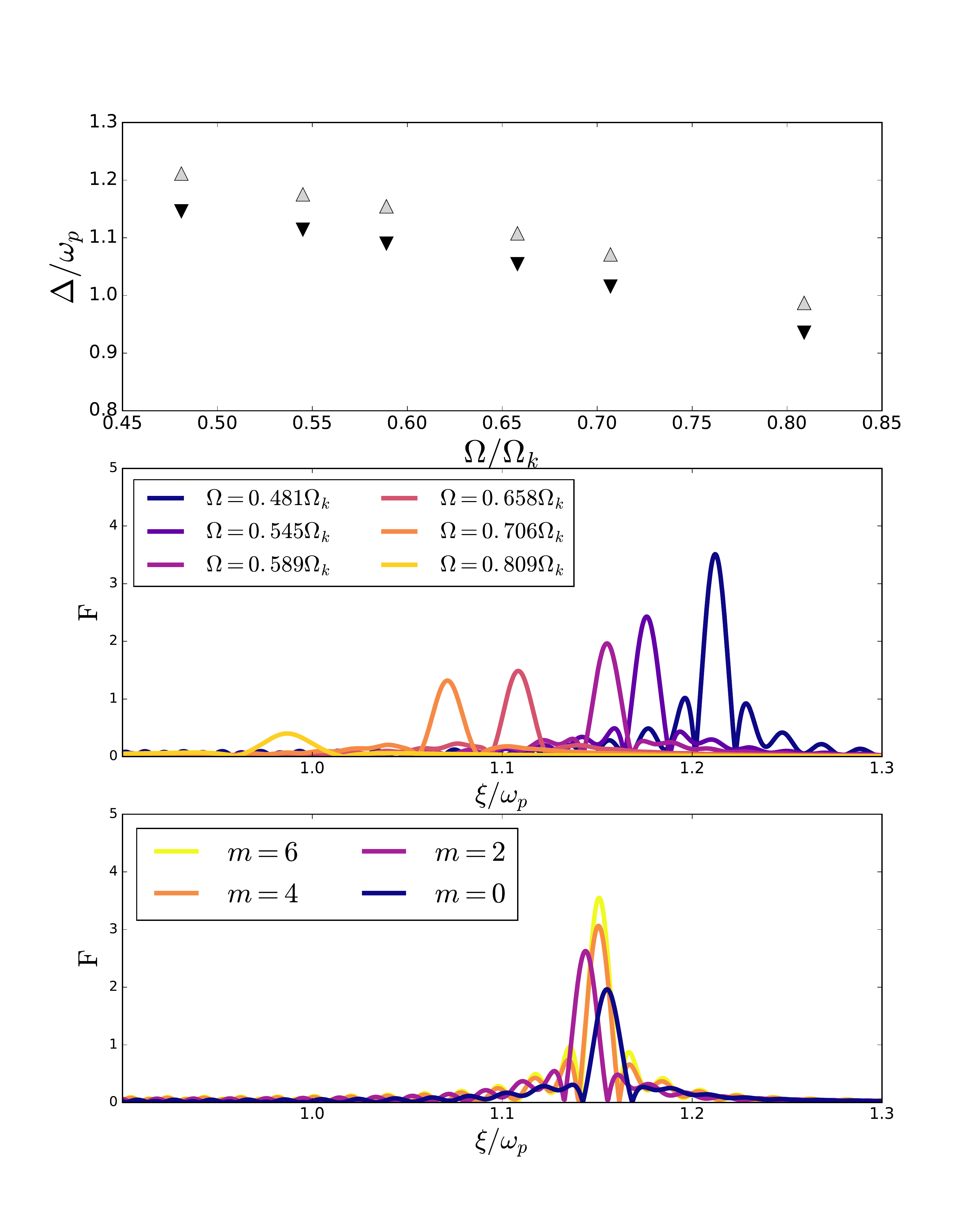}}
\caption{Upper panel : Theoretical large separation $\Delta^{\R{th}}_c/\omega_p$ (upward triangles) compared to the numerical peak's position (downward triangles) calculated for axisymmetric modes at six rotation rates. Middle panel : theoretical autocorrelations with quantum number $m = 0$, from right to left the rotation rate increases. Bottom panel : theoretical autocorrelations at $\Omega / \Omega_k = 0.589$ with $m = 0, 2, 4, 6$ and frequency $\omega = 24.41 \omega_p$.}
\label{Results}
\end{figure}

\noindent The Fourier transform of this quantity

\begin{equation*}
    F(\xi) = \frac{1}{\sqrt{2 \pi}} \int_{-\infty}^\infty \R{d}T \, \exp(-i \xi T) \, T \R{P_{\, \Omega}}(T),
\end{equation*}

\noindent is shown in the middle panel of Fig.~\ref{Results} for six rotations corresponding to the simulated chaotic spectra. This quantity is a semiclassical approximation of $C(\xi)$ (see Eq.~\ref{eq:c_of_xi}) and is closely related to the autocorrelation $R_2(\xi)$ (see Fig.~\ref{autocorrelations}). The results show that the periodic orbit theory based on the ray model indeed predicts a peak in the autocorrelation. In the middle panel of Fig.~\ref{Results}, its theoretical position $\Delta_c^{\R{th}}$ moves with rotation, as in the numerical mode computations, from $1.1458 \, \omega_p$ at $\Omega/\Omega_k = 0.481$ to $0.9254 \, \omega_p$ at $\Omega/\Omega_k = 0.809$. In the top panel, $\Delta_c^{\R{th}}$ is compared to the peak position of the numerical modes $\Delta_c$, with good agreement. %\red{We should discuss the error ? We could show the error as a function of frequency : fixed $\Omega$, 150 frequencies, shift the frequency domain.} 
The position, height and width of the peaks are completely determined by $T_0$ and $\sigma_0$. The position is found straightforwardly as $\Delta_c^{\R{th}} \approx 2 \pi / T_0$. The height and width are controlled by the ratio $\sigma_0/T_0$. Indeed, the condition $\sigma_0 \ll T_0$ is necessary to clearly distinguish the packets in the distribution of travel times, as in Fig.~\ref{distribution}. Due to the increase of $\sigma_0/T_0$ with rotation, the peak gets less visible at high rotation rates. \\

The theoretical large separation $\Delta_c^{\R{th}}$ is not only a function of the rotation rate $\Omega$, but also of the projected angular momentum $\tilde{L}_z$. In the bottom panel of Fig.~\ref{Results} we show the predicted peaks for $\tilde{L}_z = 0, 0.0819/\omega_p, 0.1638/\omega_p$ and $0.2458/\omega_p$ at $\Omega / \Omega_k = 0.589$. The shift in position is governed by the value of $T_0$ and the change in amplitude by the value of $\sigma_0$ as $\tilde{L}_z$ increases. The main difference with the axisymmetric case is a reduction of the domain of propagation, resulting in a smaller standard deviation $\sigma_0$. As the mean travel time $T_0$ between two rebounds depends on the impact of $\tilde{L}_z$ on the ray paths, it changes slightly in a non monotonic fashion. \\

%\red{The theoretical large separation $\Delta_c^{\R{th}}$ is not only a function of the rotation rate $\Omega$, but also of the projected angular momentum $\tilde{L}_z$. In the bottom panel we show the predicted peaks for  Quantitatively, we get $\Delta_c^{\R{th}} = 1.1505 \omega_p$ for $\tilde{L}_z = 0.1638 / \omega_p$, which is $0.35\%$ less than the value of $\Delta_c^{\R{th}}$ in the axisymmetric case.} \\

We have seen that the sound speed profile in the star imposes restrictions on the travel times of acoustic rays. In spite of the chaotic nature of the dynamics, the small dispersion of travel times produces order in the spectra, in the form of a peak in the autocorrelation at a value $\Delta_c^{\R{th}}$ that corresponds to the mean travel time between two points at the surface. It is fundamentally a radial phenomenon, since the strong variation of the sound speed occurs in the radial direction. Introducing the radial acoustic time at a given colatitude $\tau(\theta) = \int_0^{r_s(\theta)} \R{d}r / \tilde{c}_s$, where $r$ is the radial coordinate, the mean acoustic time between two rebounds can be estimated without using ray dynamics, as

\begin{equation}
\label{eq:average}
T^{\R{av}} = 2 \left( \frac{2}{\pi} \int_0^{\pi/2} \tau(\theta) \R{d}\theta \right),
\end{equation}

\noindent where the integral is performed over a quarter of the star because of the axial and equatorial symmetries. In the $m \neq 0$ case, the domain of integration must be adapted, since the acoustic ray cavity is reduced in size. We find that $2 \pi / T^{\R{av}}$ is a good  estimation of $\Delta_c$. For instance, at $\omega / \Omega_k = 0.589$, $2 \pi / T^{\R{av}} = 1.0996 \omega_p$, which differs from $\Delta_c$ by less than $1\%$.

%\begin{figure}[t]
%\centerline{\includegraphics[width=0.5\columnwidth]{cs_star.png}}
%\caption{Sound speed profile at $\Omega / \Omega_k = 0.706$. The dashed line is a curve of constant sound speed. \blue{Python/cs\_2D/plot\_aa.py}}
%\label{cs_71}
%\end{figure}

\subsection{Secondary peaks}
\label{secondary_peaks_2}

At every rotation there are other peaks in addition to the main peak caused by the large separation of chaotic modes. At rotations higher than $\Omega/\Omega_k = 0.589$ some of these peaks raise high above the noise level and we call them secondary peaks. The presence of small amplitude peaks in the autocorrelation can be understood as reflecting the organization of the chaotic spectra on an \'echelle diagram, illustrated in Fig.~\ref{echelle_chaos}. Indeed, the spectrum is organized in families of nearly aligned frequencies. Lets us consider first the case of two perfectly aligned families of frequencies. These two families would produce a peak in the autocorrelation at a position given by the distance between consecutive levels in the two series. This is reminiscent of the case of island modes, that show well-aligned families on the \'echelle diagram and, accordingly, many peaks are seen in the autocorrelation. In this later case, the effect is strong because the separation $\delta_\ell = \omega_{n, \ell+1} - \omega_{n, \ell}$ between consecutive families is fixed. In the case of chaotic modes, the alignments are weaker and consecutive series are not regularly spaced out. Thus, we expect the peaks to be of small amplitude. \\

The high amplitude peaks observed in the data at $\Omega/\Omega_k = 0.589$ are caused in part by two series, number 1 and 2 in Fig.~\ref{echelle_chaos}, that are nearly parallel on the \'echelle diagram. The spacing between these two series indeed corresponds to the position $\delta$ of the leftmost secondary peak in Fig.~\ref{autocorrelations}, panel c. If couples of nearly parallel series can occur occasionally, there is however no reason to expect them at every rotation rate. The presence of strong secondary peaks at other rotation rates (and also in the spectra of non axisymmetric modes) is a hint that a more generic mechanism is at play. Moreover, we know that the spectral statistics of the chaotic spectrum at $\Omega/\Omega_k = 0.706$ indicates that it is divided into independent subspectra, as explained in section \ref{sec:secondary_peaks}. In the following, we will argue that the presence of secondary peaks is a consequence of the presence of partial barriers in the phase space of the ray system which create separate independent subsets of modes. \\

It is known that transport properties in chaotic phase space can have a significant impact on the wave system spectra \citep{bohigas_manifestations_1993}. The transport of trajectories from a subregion A to another subregion B can be affected by the presence of partial barriers. These are curves through which classically trajectories can flow, but with a much smaller flux than in other parts of phase space. Thus, contrary to KAM tori that act as complete barriers, ergodic trajectories are able to cross partial barriers after a sufficiently long time. Such partial barriers are typically created when the system is being perturbed, through the destabilization of island orbits. In virtue of the Poincar\'e-Birkhoff theorem \citep{ott_chaos_1993}, resonant tori are destroyed by the perturbation and generate new structures in phase space : a new (smaller) island chain around an elliptic central orbit along with an unstable periodic orbit. The unstable orbits created through this process are known to be the source of partial barriers that trap ergodic trajectories \citep{shim_whispering_2011,bohigas_manifestations_1993}. \\

\begin{figure}[!htbp]
\centerline{\includegraphics[width=1\columnwidth]{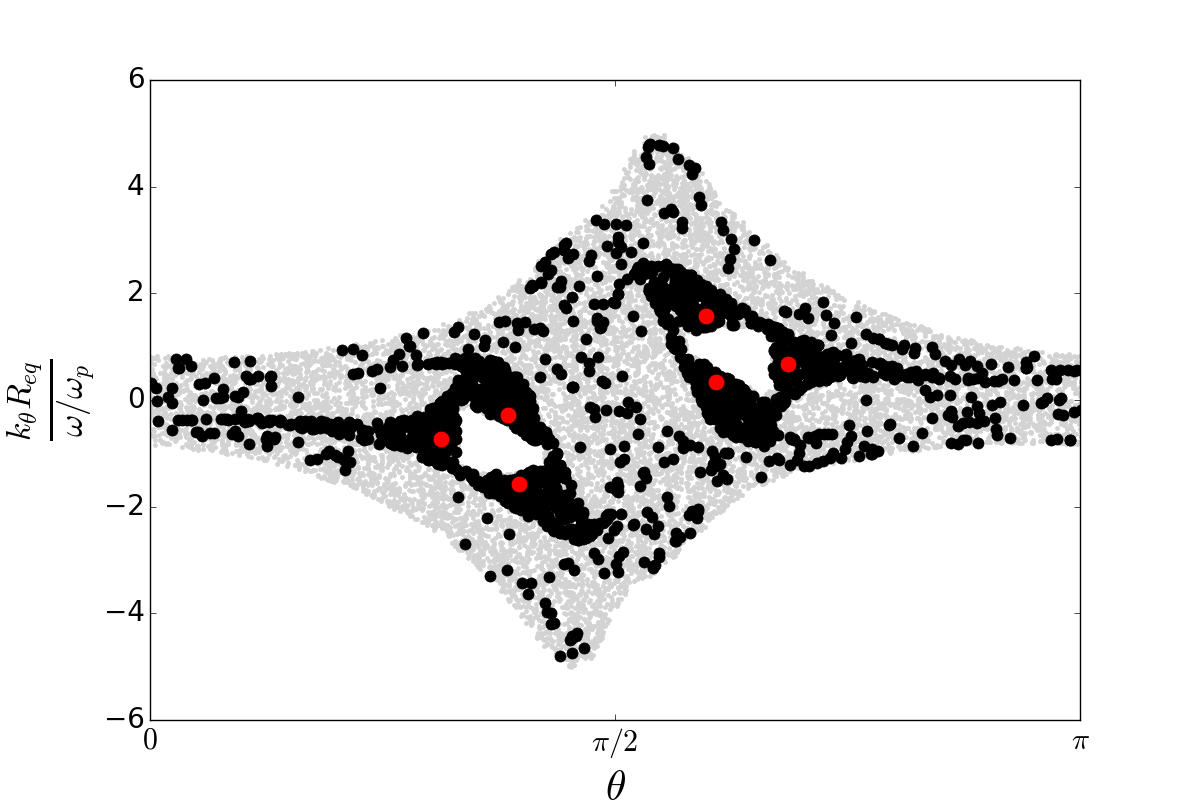}}
\caption{Snapshot of the evolution of a bundle of trajectories as they intersect the PSS, represented by black dots, at $\Omega / \Omega_k = 0.706$. The phase space zones where black dots are dense correspond to regions enclosed by partial barriers. The trajectories are initially in the neighborhood of the two main islands. One of the central periodic orbits of these zones is indicated by colored dots (large gray dots).}
\label{barriers}
\end{figure}

Acoustic ray simulations show that two 6-period unstable orbit revolves around the main 2-period island chain at $\Omega/\Omega_k = 0.589$. At $\Omega/\Omega_k = 0.706$, we also find a periodic orbit revolving around the island and strongly suspect the presence of a second one. Moreover, by choosing trajectories with initial conditions around the main island and evolving the system forward in time, we see clearly the contours of the partial barrier reveal themselves (see Fig.~\ref{barriers}). For the frequencies considered in our dataset, the partial barriers may act as complete barriers and isolate some modes, as observed in other systems \citep{shim_whispering_2011}. Quantitative estimates of the area and outgoing flux have been given in \citet{evano_correlations_2019}, showing that the barrier grows in size from $\Omega / \Omega_k = 0.589$ to $\Omega / \Omega_k = 0.706$ and that, in parallel, it takes longer to go through the partial barrier at $\Omega / \Omega_k = 0.706$. Thus, the trapping of chaotic trajectories around stable islands seems to be the cause of the additional peaks seen in the autocorrelations. Moreover, as the trapped trajectories revolve around a 6-periodic orbit (see Fig.~\ref{barriers}), and assuming the modes to quantize like island modes, one can expect that the secondary peaks will be located approximately at $\Delta_c/3$. As seen in  Fig.~\ref{autocorrelations}, this idea is consistent with the data at $\omega / \omega_k = 0.706$. However it gives only a rough estimate of the peak position at other rotation rates and thus this issue should be analyzed in more depth. %However, no quantitative prediction can be made regarding either the position or amplitude of the additional peaks \red{(acoustic time along the two periodic orbits ?)}.
The specific rotations where the trapping of trajectories will be efficient can be anticipated only through precise numerical simulations of the ray model. \\

%The ratio of the region enclosed by partial barriers to the complete chaotic zone area is $12\%$ at $\Omega / \Omega_k = 0.589$ and $25 \%$ at $\Omega / \Omega_k = 0.706$. To quantify the outgoing flux we computed the escape rate $\lambda$ of the trajectories off the partial barrier zone. This gives $\lambda(\Omega / \Omega_k = 0.589) = 0.17$ and $\lambda(\Omega / \Omega_k = 0.706) = 0.06$.

\subsection{Amplitude distribution of chaotic modes}
\label{amplitude}

Berry \citet{} proposed that the eigenstates of a classically chaotic system, like quantum billiards, could be modeled by a superposition of random plane waves. Indeed this picture holds locally, in the high frequency regime, where the mode is a superposition of rays of the same magnitude $|k|$, but coming from all possibles directions. It is easy to test this idea, as done e. g. in \citet{oconnor_properties_1987}, by adding multiple time-harmonic solutions of the Helmholtz equation $\nabla^2 \psi + k^2 \psi = 0$. The waves are of the form $\psi_n = a_n \, \cos(\B{k_n}(\alpha_n) \cdot \B{x} + \xi_n)$, where the amplitude $a_n$, the wave vector orientation $\alpha_n$ and the phase shift $\xi_n$ are random variables but the magnitude $|k|$ of the wave vector is fixed. The resulting modes indeed exhibit the random ridges characteristic of chaotic modes. \\

\begin{figure}[!htbp]
\centerline{\includegraphics[width=1\columnwidth]{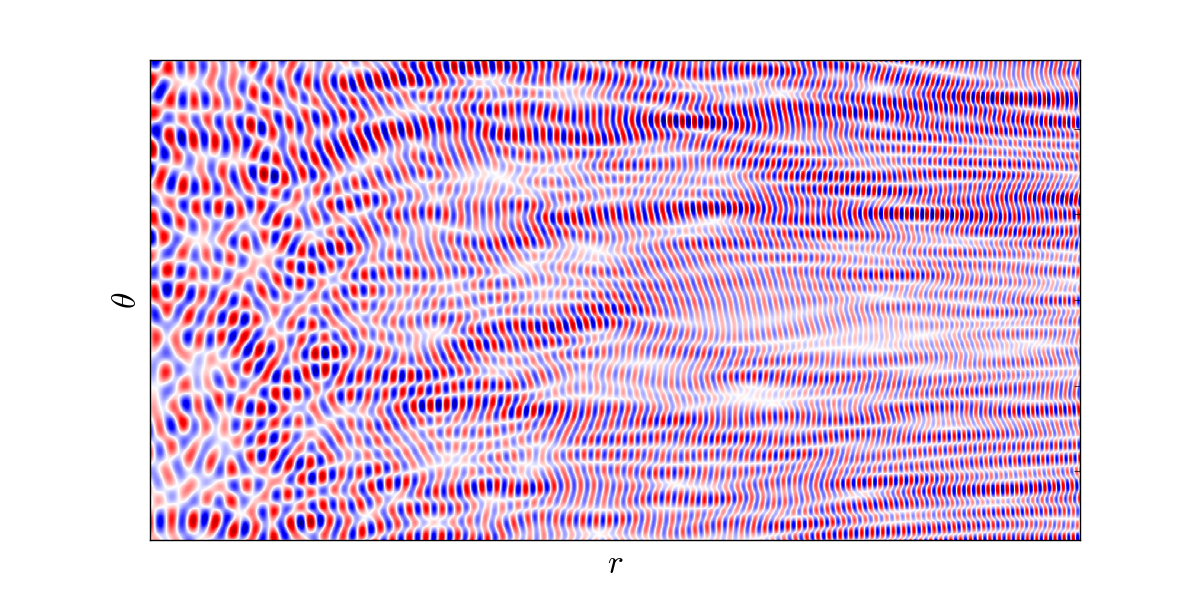}}
\caption{Superposition of random stationary waves reproducing the kind of amplitude patterns seen in the chaotic p-modes of the polytropic stellar model (see text).}
\label{random_waves}
\end{figure}

\noindent Contrary to quantum billiards, the wavevector magnitude varies strongly within a star, as it is proportional to the inverse of the sound speed. Moreover, each incoming ray approaching the surface is almost aligned in the radial direction ($k_r \gg k_{\theta}$). Thus, the hypothesis that intersecting rays come with a large variety of possible orientations is not valid near the surface. We modified slightly the random wave model to incorporate such a behavior. For simplicity, we treat the x axis as a radial direction. Then we impose a radial increase of the magnitude $|k|$ and that all waves end up aligned in the radial direction as $x \rightarrow \infty$. An example of mode produced in this way is shown in Fig.~\ref{random_waves}. The mode exhibits random ridges in the center but the outer part is more structured, with nodes regularly spaced radially and irregular in the transverse direction as in the star.

\subsection{Spectrum organization}
\label{families}

In the previous sections, we found that the chaotic spectrum can be described as a set of series of modes, where a series corresponds to modes separated by approximately $\Delta_c$ and having similar amplitude distribution. Sixteen series have been effectively identified at the rotation rate $\Omega/\Omega_k = 0.589$ and in the frequency range, $25.60\,\omega_p < \omega < 33.54\,\omega_p$. We also found that whispering gallery modes are present at the low-frequency end of some chaotic series, and that island mode series appears as high-frequency extensions of some other chaotic series. In this section, we use semiclassical arguments to investigate the origin of the different chaotic series and their link with the whispering gallery and island mode spectra.

\begin{figure}[!htbp]
\centerline{\includegraphics[width=1\columnwidth]{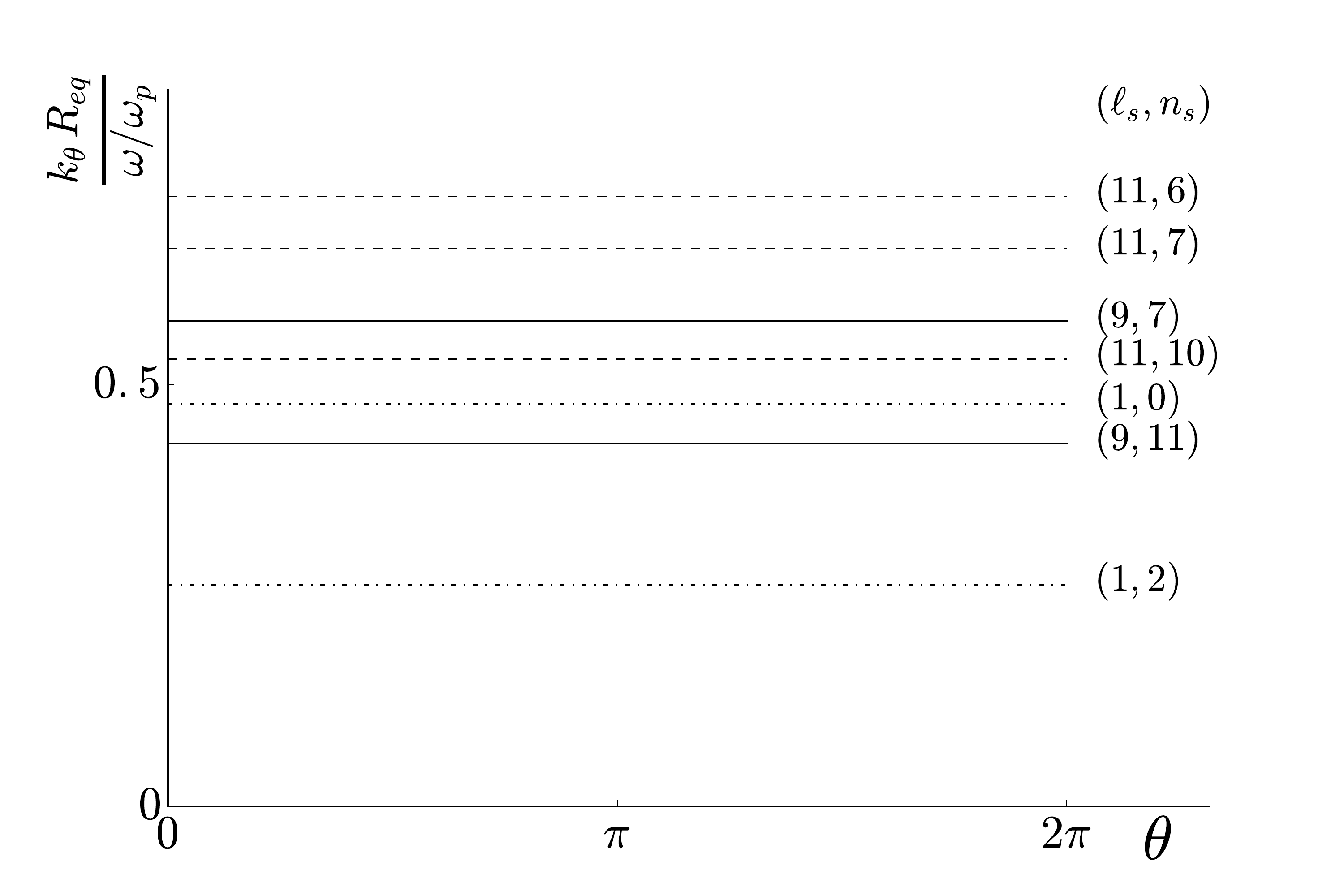}}
\caption{Intersection of a few tori with the PSS at $\Omega / \Omega_k = 0$. Only the part of the PSS where $k_\theta >0$ is shown because of the symmetry with respect to the $k_\theta = 0$ axis at zero rotation. The tori correspond to modes of degree $\ell_s = 1$, $\ell_s = 9$ or $\ell_s = 11$.}
\label{fig:tori}
\end{figure}

We first consider a non rotating star because in this case the modes can be precisely located on the PSS. As the system is integrable, the phase space is foliated with $N_d$-dimensional tori, where $N_d$ is the number of degrees of freedom. Modes are then constructed on some of the tori, the ones specified by the Einstein-Brillouin-Keller quantization rules. The quantization of the norm of the angular momentum $L$, the invariant associated with the spherical symmetry, reads $L = \pm (\ell_s + 1/2)$ \citep{gough_linear_1993} where $\ell_s$ is the degree of the mode, with the index $s$ denoting the "spherical" case. For the frequency-scaled coordinates used here, we rather use the invariant $\tilde{L} = L/ \omega = \pm (\ell_s + 1/2) / \omega_{n_s, \ell_s, m}$, where $n_s$ and $m$ are the radial order and the azimuthal order of the mode. In the axisymmetric case, $L_z = 0$, the tori imprint the PSS on horizontal lines $\tilde{k}_\theta = \pm \tilde{L}$ \citep{lignieres_asymptotic_2009}. On Fig.~\ref{fig:tori}, the identification of a few mode-carrying tori is displayed. It shows how the position of the tori along the vertical axis depends on $\ell_s$ and $n_s$. For fixed radial order $n_s$ and increasing $\ell_s$, the tori indeed move toward higher $\tilde{L}$ or $\tilde{k}_\theta$ values. Similarly, for fixed $\ell_s$ and increasing $n_s$, the tori approach the $\tilde{k}_\theta = 0$ axis. In particular, the $\ell_s = 0$ or $\ell_s = 1$ modes are already close to the central axis for the smallest radial order.

%$\tilde{L} = L/\omega = \pm (\ell_s + 1/2)/ \omega_{n_s,\ell_s,m}$, where $n_s$ and $m$ are the radial order and azimuthal number of the mode. In the axisymmetric case, $L_z =0$, the tori imprint the PSS on horizontal lines $\tilde{k}_{\theta} = \pm \tilde{L}$ \citep{lignieres_asymptotic_2009}. On Fig. \ref{fig:tori}, the identification of a few mode-carrying tori is displayed. It shows how the position of the tori along the vertical axis depends on $\ell_s$ and $n_s$. For fixed radial order $n_s$ and increasing $\ell_s$, the tori indeed move towards higher $\tilde{L}$ or $\titde{k}_{\theta}$ values. Similarly, for fixed $\ell_s$ and increasing $n_s$, the tori approaches the  $\tilde{k}_{\theta} = 0$ axis. In particular, the $\ell_s = 0$ or $\ell_s =1$ modes, are already close to the central axis for the smallest radial order.

When rotation comes into play, the location of the tori at $\Omega = 0$ has strong consequences on their fate. From the evolution of the  PSS we know that the phase space becomes rapidly dominated  by three main structures : the 2-period island chain at low $\tilde{k}_{\theta}$, the large chaotic zone at low and intermediate $\tilde{k}_{\theta}$, and at large $\tilde{k}_{\theta}$, the region of surviving KAM tori corresponding to whispering gallery trajectories. In this context, high-$\tilde{L}$ tori at $\Omega = 0$ will transform into structures of the surviving KAM tori region, whereas low-$\tilde{L}$ tori will be destroyed as the 2-period island chain and the chaotic zone surrounding it emerge. We thus expect high-$\tilde{L}$ modes to become whispering gallery modes, while low-$\tilde{L}$ modes should evolve toward chaotic or island modes.\\

We can use this phenomenology to predict the fate of a sequence of modes having a fixed degree $\ell_s$ and variable radial orders $n_s$. In such a series, $\tilde{L}$ has a maximum value for $n_s=1$ and it decreases toward zero as $n_s$ increases. Generically, we thus expect that, as rotation increases, the low-$n_s$ (high-$\tilde{L}$) modes become whispering gallery modes, the intermediate-$n_s$ modes become chaotic modes, and the high-$n_s$ (low-$\tilde{L}$) modes become island modes.
Combining this picture with our observation that some series of chaotic modes show island modes at their high-frequency end or whispering gallery modes at their low-frequency end, we are led to interpret the chaotic series together with their island modes and whispering gallery modes extensions as the remnants of series of given $\ell_s$ at zero rotation.

At $\Omega/\Omega_k = 0.589$, we could indeed relate two chaotic series (the series $1$ and $2$) with the $\ell=4$ and the $\ell=5$ odd-parity island mode series, respectively. Following this interpretation and using the formulas that link $\ell$ to $\ell_s$ \citep{reese_regular_2008, pasek_regular_2012} we can attribute the $\ell_s=9$ value to series $1$ and the $\ell_s=11$ value to series $2$. In principle, the $\ell_s$ value of the other chaotic series shown on Fig.~\ref{echelle_2} could also be determined, either by following them to higher frequency up to the island mode transition or by looking for a whispering gallery mode and its $\ell_s$ value at low frequency.\\

The generic case of a $\ell_s$ series containing the three type of modes only holds for high enough $\ell_s$. The computed PSS indeed show that below a specific $\tilde{k}_{\theta}$ that depends on the rotation rate, all trajectories are either chaotic or within an island chain. We thus expect that, below some critical $\ell_s$ that depends on the rotation rate, all modes in the series are either of the chaotic or island type. Moreover, from numerical studies \citep{lignieres_acoustic_2006, reese_regular_2008,pasek_regular_2012} where $\ell_s =\{0,1,2,3\}$ modes have been carefully followed with rotation, we know that, for these lowest degree, modes of all orders behave as island modes. This is coherent with the fact that their initial $\tilde{L}$ are all close to the $\tilde{k}_{\theta} =0$ axis.\\

To summarize, we argued that the chaotic mode spectrum of rapidly rotating stars is organized in series that can be traced back to the fixed $\ell_s$ series of the non-rotating star. While supported by the analysis of our numerical results, this idea needs to be further tested at other rotations and in other frequency domains. It is also important to stress that without the $\Delta_c$ organization of the chaotic spectrum, the chaotic series would not exist and we could not attribute  $\ell_s$ value to them. Indeed, tracing back chaotic modes to their integrable counterpart is not possible for typical mixed systems like quantum billiards. Another property that help recognize the link with a high-frequency island mode series is that $\Delta_i$ is very close to $\Delta_c$.

\subsection{Mode identification : chaos versus islands}
\label{sec:chaos_vs_islands}

In the numerically computed spectra, the large separations of island modes and chaotic modes are found to be very close to one another (see table~\ref{table:2}). The asymptotic theory discussed above gives a natural explanation for this apparent coincidence. Indeed we established that the large separation of chaotic modes is defined asymptotically by $\Delta_c^{\R{th}} = 2 \pi / T_0$. On the other hand, the large separation of island modes in the asymptotic regime is related to the acoustic time along the central  periodic orbit $\gamma$ by \citep{pasek_regular_2012} $\Delta_i  = 2 \delta_n$ with $\delta_n =  2 \pi / \oint_{\gamma} (\R{d}s/\tilde{c}_s)$. %The factor $2$ in the last expression comes from the fact that the integration runs over a complete orbit period, that is twice the distance between two surface bouncing points. 
The closeness of the two peaks is thus due to the closeness of the mean travel time of a chaotic trajectory and the travel time along the central path $\gamma$. For the same reason that all chaotic trajectories have almost the same travel time, the acoustic time between two rebounds along the central orbit of the island has to be very close to $T_0$.

\begin{table}
\caption{Comparison of the large separation of island modes $\Delta_i$ and chaotic modes $\Delta_c$, obtained from the simulated spectra at six rotations}
\label{table:2}
\centering
\begin{tabular}{c c c c}
\hline\hline
$\Omega / \Omega_k$ & $\Delta_i$ & $\Delta_c$ & $|\Delta_i - \Delta_c| / \Delta_i$ \\
\hline
    $0.481$ & $1.1644 \, \omega_p$ & $1.1458 \, \omega_p$ & $ 1.60 \, \%$ \\
    $0.545$ & $1.1288 \, \omega_p$ & $1.0907 \, \omega_p$ & $ 3.38 \, \%$ \\
    $0.589$ & $1.1021 \, \omega_p$ & $1.1132 \, \omega_p$ & $ 1.01 \, \%$ \\
    $0.658$ & $1.0543 \, \omega_p$ & $1.0543 \, \omega_p$ & $ 0.00 \, \%$ \\
    $0.706$ & $1.0155 \, \omega_p$ & $1.0169 \, \omega_p$ & $ 0.14 \, \%$ \\
    $0.809$ & $0.9345 \, \omega_p$ & $0.9355 \, \omega_p$ & $  0.10 \, \%$ \\
\hline
\end{tabular}
\end{table}

Despite the proximity of the two peaks, it may be possible to tell them apart by combining odd and even spectra. In this case,  the autocorrelation of the island mode spectra shows not only a peak at the large separation but also at half the large separation. This is due to the fact that the island modes are built around a central orbit which is self-retracing, i.e. during a complete period it goes twice through the same points in $q$ with opposite momenta. For such orbit, the semiclassical quantization for an even or odd spectrum uses a twice shorter orbit than for the full spectrum.  In contrast, the odd and even spectra of chaotic modes are built on generic orbits with no such property, and are completely independent. Hence, the autocorrelation of the full chaotic spectrum at a given rotation, with both parities, does not show a peak at half separation.

\section{Discussion and conclusion}
 %\red{Another potentially crucial development is the possibility to constrain the rotation rate from the low-frequency part of the spectrum (cit : Balona et al. 2011, Van Reeth 2015, Christophe et al. 2018, Saio et al. 2018). Having a reliable rotation would greatly improve our chance to interpret the acoustic part of the spectrum. TESS bright stars.} \\
 
In this paper, we have computed high frequency p-modes in stellar polytropic models for rotation rates between $\Omega / \Omega_k = 0.48$ and $\Omega / \Omega_k = 0.81$. Following the methodology of section \ref{sec:mode_identification}, we have then identified chaotic modes and built a dataset of chaotic frequencies. As expected, the nearest-neighbor statistics of the chaotic spectra for most rotations follow the Wigner-Dyson distribution, a well-known generic property of wave chaos systems. The frequency autocorrelations of the chaotic spectra have been computed. All of them exhibit peaks above the noise level. The presence of peaks in the frequency autocorrelation is clearly not generic in wave chaos systems.\\

Our analysis shows that chaotic modes are organized in series. The frequency difference between consecutive modes being approximately constant and of similar value across all series. By displaying chaotic mode frequencies on \'echelle diagrams, we showed that this weakly varying frequency interval can be interpreted as a pseudo large separation. We speak about pseudo large separation because contrary
to modes in the non-rotating case or to island modes, the interval is slightly irregular and would remain so asymptotically.

The pseudo large separation is responsible for the presence of the so-called main peak in the frequency autocorrelations and we explained it using semiclassical methods. The ray dynamics is indeed peculiar, as the sound speed is strongly inhomogeneous along the radius of the star. We characterized the impact of the sound speed profile on the ray dynamics through two variables, $\sigma_0$ and $T_0$ (see Eq.~\ref{podist}). These two quantities correspond respectively to the mean value and standard deviation of the one-chord travel time distribution. Using the formalism of quantum chaos, we then wrote a semiclassical expression of the autocorrelation and showed that knowing $\sigma_0$ and $T_0$ is enough to recover the position of the main peak in the numerically computed spectra. This asymptotic analysis also explains the decrease of the peak height as rotation increases, equivalent to a loss of regularity of chaotic modes. There are other peaks in the autocorrelations, which vary in a less predictable way as a function of rotation. We propose that they are created by the presence of phase space structures that develop around the stable island chains, called partial barriers. \\

The large separation is expressed as $\Delta_c \sim 2 \pi / T_0$. Since chaotic trajectories cover the entire meridional plane, at least in the asymmetric case, we found that $T_0$ can be estimated approximately, without using ray tracing, by computing the average acoustic time over the meridional plane (see Eq.~\ref{eq:average}). \\

As explained in section \ref{sec:chaos_vs_islands}, the small variance of acoustic travel times ($\sigma_0/T_0 \ll 1$) implies the observed quasi-degeneracy of $\Delta_i$ and $\Delta_c$. Thus, we expect rapidly rotating stars to be characterized by a unique large separation $\Delta \sim \Delta_i \sim \Delta_c$. 
%Concerning the autocorrelations, our results suggest that chaotic modes should produce a signal at the large separation, rather than adding noise as was previously thought. 
Autocorrelation peaks at the large separation detected in $\delta$ scuti stars \citep{garcia_hernandez_observational_2015} could be produced not only by island modes, as previously thought, but also in part by chaotic modes. This would be important for stars that rotate rapidly enough to harbor a significant number of chaotic modes. A way to distinguish the contribution of island modes is to look for a peak at half the large separation as our analysis indicates that it is due to island modes only. \\
%It must be noted, however, that even chaotic modes are less regular than odd chaotic modes. This is due to the stronger effect of avoided crossings on even mode frequencies. \\

To go further in the comparison with observed spectra, one should construct a database of low-frequency synthetic spectra as in \citet{reese_frequency_2017} but with full mode identification and with a higher sampling in rotation rate. We expect the asymptotic properties described in the present paper to guide the identification of chaotic modes even at low frequency, as it was the case for island modes. Having fully identified synthetic spectra then would help to identify modes in real data.

Calculations of mode visibilities in rapid rotators were performed first in \citet{lignieres_asymptotic_2009} and later in \citet{reese_mode_2013} taking more effects into account such as gravity darkening. These calculations showed that the surface structure of chaotic modes should allow them to be visible, especially at high rotation rates. The most direct proof of the occurrence of wave chaos in stars would be to identify a large set of observed chaotic mode frequencies and find that they follow closely the Wigner-Dyson surmise. This is a very difficult task, however a couple of observations might reduce the difficulty by a small amount. First, the spectra of many stars with various rotation rates can be aggregated, as we did in Fig.~\ref{spacings} to construct the nearest neighbor distribution. Also, choosing very fast rotators may help, since chaotic modes are expected to be more present at very high rotation rates. Finally, a pole-on configuration can be helpful as avoiding $m \neq 0$ sub-spectra would simplify mode identification.  

\begin{acknowledgements}
We  thank CALMIP (“CALcul en MIdi-Pyr\'en\'ees”) for the use of their supercomputer. We used the code Top developed by D. Reese and made user-friendly by B. Putigny. We thank ISSI (“International Space Science Institute”) through the SoFAR (“Seismology of Fast Rotating Stars”) program for their support.
\end{acknowledgements}

\appendix

\section{Trace formula}
\label{trace_formula}

In this Appendix we will adapt the semiclassical formalism used in quantum mechanics to derive a trace formula for chaotic modes in rotating stars. The general derivation follows the original one due to Gutzwiller, detailed e.g. in \citep{gutzwiller_chaos_1990,chaosbook,ott_chaos_1993}.

\subsection{The Hamiltonian system}

We denote the canonically conjugate variables $(\B{p}, \B{q})$, as is usual in textbooks on Hamiltonian mechanics. In the present subsection, we will consider the system to be one-dimensional to avoid carrying indices in the notation. The ray dynamics is governed by the following Hamiltonian :

\begin{equation}
H = \sqrt{c_s^2 k^2 + \omega_c^2} = \omega,
\end{equation}

\noindent where the wave vector plays the role of momentum : $p \equiv k$.

\subsubsection{Hamilton's principal function}
\label{sec:HPF}

We now express $p$ as a function of $\dot{q}$ :

\begin{equation}
\frac{\partial H}{\partial p} = \dot{q} = 2 c_s^2 p / \omega \implies p = \frac{\dot{q} \omega}{2 c_s^2}.
\end{equation}

\noindent The ray system can thus be seen as analogous to a mechanical system with a varying mass $\omega/c_s^2$. Then, the Lagrangian $L$ is obtained through the usual Legendre transform :

\begin{equation}
\label{eq:lagrangian}
L = \dot{q} p - H = \frac{\dot{q}^2 \omega}{2 c_s^2} - \omega.
\end{equation}

\noindent Hamilton's principal function, denoted $R$, is defined as the time integral of the Lagrangian. Its computation involves following a trajectory from $(q', t')$ to $(q, t)$. For short times $\delta t$, it gives :

\begin{equation}
\label{eq:R(dt)}
R(q, q', \delta t) = \frac{(q-q')^2 \omega}{2 c_s^2 \delta t} - \omega\; \delta t,
\end{equation}

\noindent using $\dot{q} = (q-q') / \delta t$. 

\subsubsection{Action integral}

The action integral $S$ is defined from $R$ as $S(q,q',\omega) = R(q,q',t) + \omega t = \int_0^{t} (L+H) \, \R{d}t$. From Eq.~\ref{eq:lagrangian} we have :

\begin{equation}
\begin{aligned}
S(q(t), q'(0)) &= \int_0^{t} \frac{\dot{q}^2 \omega}{2 c_s^2} \, \R{d}t \\
                             &= \int_{q'(0)}^{q(t)} \frac{\dot{q}^2 \omega}{2 c_s^2} \, \frac{\R{d}q}{\dot{q}} \\
                             &= \int_{q'(0)}^{q(t)} \frac{\dot{q} \omega}{2 c_s^2} \, \R{d}q = \int_{q'(0)}^{q(t)} p \, \R{d}q,
\end{aligned}
\end{equation}

\noindent which is the well know expression of the action. Finally, the eikonal equation Eq.~\ref{eq:eikonal} gives $\tilde{k} = p/\omega = 1/\tilde{c}_s$. Thus the action can be written in terms of the acoustic time :

\begin{equation}
S(q,q',\omega) = \omega \int_{q'}^{q}\frac{\R{d}s}{\tilde{c}_s},
\end{equation}

\noindent where $s$ is the curvilinear coordinate along the ray path.

\subsection{WKB approximation for the semiclassical propagator}

We first derive the expression for the semiclassical propagator, adapting the method in  \citep{chaosbook} for quantum systems to our star model, which has two degrees of freedom and a four-dimensional phase space.
Let $\Lambda^{-1}$ be a small dimensionless parameter. In the wave equation, Eq.~\ref{helmoltz}, we insert the WKB ansatz $\Psi(q, t) = A \, e^{i \Lambda \phi(q, t)}$ leading to \citep{gough_linear_1993} :

\begin{align}
(\partial_t \phi)^2 - \frac{1}{\Lambda^2} (\omega_c^2 + c_s^2 k^2) = 0 \label{eq:phase}, \\
\Lambda \partial_t \phi \pm (\omega_c^2 + c_s^2 k^2)^{1/2} = 0.
\end{align}

\noindent Eq.~\ref{eq:phase} can be recognized as the Hamilton-Jacobi equation $ \partial R / \partial t = \pm H$. Thus, in the limit of small wavelengths, the phase is simply Hamilton's principal function : $\phi = R$ or $\phi = -R$ (associated to the Hamiltonian $-H$). The two possible phases lead to two terms in the propagator, with a projector on each subspace $P_1$ and $P_2$ satisfying $P_1+P_2=I$.\\

\noindent Additionally, the substitution $\dot{\phi} = -H = -\omega$ yields :

\begin{equation}
\label{eq:continuity}
\frac{\partial}{\partial t} (A^2) + \frac{c_s^2}{\omega} \frac{\partial}{\partial q} \left(A^2 \frac{\partial R}{\partial q} \right) = 0.
\end{equation}

\noindent Introducing the density $\rho = A^2$ and velocity $v = 1/m \, (\partial R / \partial q)$, with $m = \omega/c_s^2$ (from the mechanical analogy introduced in section \ref{sec:HPF}), Eq.~\ref{eq:continuity} appears as a continuity equation. It follows that the ray amplitude $A(q, t)$ can be interpreted as the square root of the density of nearby trajectories. The evolution of this density from $q'$ to $q$ (variation of volume in coordinate space of a swarm of trajectories) is quantified through the Jacobian determinant $\det \left(\frac{\partial q'}{\partial q}\right) $.

%\noindent Let us consider a swarm of trajectories starting at $q' = q(0)$. At time $t>0$, the \red{same} number of trajectories \red{can lead to a different occupied coordinate space volume. This can be written}

%\begin{equation}
%\rho(q, t) = \det \left(\frac{\partial q'}{\partial q}\right) \rho(q', 0)
%\end{equation}

%\noindent where the ratio of the volumes has been expressed \red{through the Jacobian} determinant. 

Having now both the phase and amplitude, we obtain the semiclassical wave function

\begin{equation}
\label{eq:classical_wave}
\begin{aligned}
\Psi_{sc}(q, t)& = A_1\int dq' \left| \det \frac{\partial q'}{\partial q}\right|^{1/2} e^{i R(q,q',t)} \Psi_{sc}(q', 0) \\
&+ A_2\int dq' \left| \det \frac{\partial q'}{\partial q}\right|^{1/2} e^{-i R(q,q',t)} \Psi_{sc}(q', 0),
\end{aligned}
\end{equation}

where $A_1$ and $A_2$ are the projection of the intial wave function on the two subspaces. This is valid for short time, i.e. when only one classical trajectory connects $q'$ to $q$ in time $t$. For longer times, several trajectories which we label by $j$ connect the two points, and the formula becomes:

\begin{equation}
\label{eq:final}
\begin{aligned}
\Psi_{sc}(q, t) &=  A_1 \int dq' \sum_j \left| \det \frac{\partial q'}{\partial q}\right|^{1/2}  e^{i R_j(q,q',t) - i \kappa_j \pi/2} \Psi_{sc}(q', 0) \\
&+ A_2 \int dq' \sum_j \left| \det \frac{\partial q'}{\partial q}\right|^{1/2}  e^{-i R_j(q,q',t) - i \kappa_j \pi/2}  \Psi_{sc}(q', 0),
\end{aligned}
\end{equation}

where the topological index $\kappa$ is added to account for the phase shift at points where the amplitude becomes singular, such as caustics.

The propagator $K$ is defined by

\begin{equation}
\label{semic}
\psi(q,t) = \int dq' K(q, q', t) \, \psi(q', 0).
\end{equation}

\noindent The propagator is the time-dependent Green's function, it is thus solution of the wave equation with the initial condition $\mbox{lim}_{t\rightarrow 0} K(q, q', t) = \delta(q-q')$. 
%Identifying with Eq.~\ref{eq:final} (or, alternatively, starting with a WKB form for the propagator and inserting into Eq.~\ref{helmoltz}), and using that $\frac{\partial p'}{\partial q''} = -\frac{\partial^2 R}{\partial q'' \partial q'}$, we obtain the semiclassical expression:
Again, we will assume that, for short times $\delta t$, the semiclassical propagator is of the form (for the first term, the computation is similar for the second term)

\begin{align}
K_{sc}(q, q',  \delta t) = A(q, q', \delta t) \, e^{i R(q, q', \delta t)}.
\end{align}

\noindent Neglecting the second term $\omega \; \delta t$ in \ref{eq:R(dt)} gives

\begin{equation}
K_{sc}(q,q',\delta t) = A(q, q', \delta t) e^{i \frac{\omega}{2 c_s^2 \delta_t}(q-q')^2}.
\end{equation}

\noindent If we impose $A(q, q', \delta t) = (\frac{\omega}{2 \pi i c_s^2 \delta t})$, then the previous expression is a 2-dimensional Gaussian of width $\sigma = (\delta t \, c_s^2 / \omega)^{1/2}$. In the limit $\delta t = 0$ the condition $K(q'',q', t=0) = \delta(q-q')$ is satisfied. Using again the substitution $m = \omega/c_s^2$ for clarity

\begin{equation}
K_{sc}(q, q', \delta t) = \left( \frac{m}{2 \pi i \delta t} \right) e^{i R(q,q', \delta t)},
\end{equation}

using the fact that $m/\delta t = \det (-\frac{\partial p'}{\partial q})=\det (-\frac{\partial^2 R}{\partial q \partial q'})$ the expression becomes

\begin{equation}
\label{eq:sc_propagator}
K_{sc}(q ,q',t) = (2 \pi i)^{-1} \left|\det \frac{\partial^2 R}{\partial q \partial q'} \right|^{1/2} e^{i R_j(q,q',t) }.
\end{equation}

This short time expression with the correct limit at $t\rightarrow 0$ can be extended to longer times using Eq.~\ref{eq:final} and the fact that $\det \frac{\partial p'}{\partial q} \det \frac{\partial q}{\partial q''} = \det \frac{\partial p'}{\partial q''}$. This gives: 

\begin{equation}
\label{eq:sc_propagator}
\begin{aligned}
K_{sc}(q ,q',t) &= A_1\sum_j (2 \pi i)^{-1} \left|\det \frac{\partial^2 R}{\partial q \partial q'} \right|^{1/2} e^{i R_j(q,q',t) - i \kappa_j \pi/2} \\
&+ A_2 \sum_j (2 \pi i)^{-1} \left|\det \frac{\partial^2 R}{\partial q \partial q'} \right|^{1/2} e^{-i R_j(q,q',t) - i \kappa_j \pi/2},
\end{aligned}
\end{equation}

%\noindent where the mass $m$ is hidden in the determinant through $\partial R / \partial q = p \approx m \dot{q}/ \delta t$. Finally\
where the sum is over all classical trajectories labelled by $j$ from $q'$ to $q$ in time $t$. %The term $(2 \pi i)^{-1/2}$ corresponds to the normalization to get the correct behaviour for $t \rightarrow 0$.

\subsection{Green's function}

To derive the trace formula, the usual procedure necessitates in quantum mechanical systems to go from the propagator to the energy-dependent Green function $G(q,q',E)$, which is related to the propagator through the Fourier transform

\begin{equation}
G (q, q', E) = \frac{1}{i \hbar} \int_0^\infty K(q, q', t) \, e^{\tfrac{i}{\hbar}Et +\tfrac{\epsilon t}{\hbar}} \R{d}t,
\end{equation}

\noindent where $\epsilon$ is a small positive number which makes the integral convergent and goes to zero eventually. Then one uses the fact that $G(q,q',E)$ can be expanded on a basis of eigenvector $\{ \phi_j \}$ of the Hamiltonian with eigenvalues $E_n$ as

\begin{equation}
G(q,q',E) = \sum_n \frac{\phi_n^*(q)\phi_n(q')}{E-E_n + i \epsilon}.
\end{equation}

\noindent In the semiclassical approximation this leads to the following equality for $E/\hbar \rightarrow \infty$

\begin{equation}
\label{eq:green_equality}
\sum_n \frac{\phi_n^*(q)\phi_n(q')}{E-E_n+ i \epsilon} = \frac{1}{i \hbar} \int_0^\infty K_{sc}(q, q', t') \, e^{\tfrac{i}{\hbar}Et} \R{d}t,
\end{equation}

\noindent which is valid at first order in $\hbar$ (acoustic case : first order in $\Lambda^{-1}$). One can already forecast the trace formula from this equality, as it involves both the eigenenergies of the quantum system on the left-hand side and classical quantities in the right-hand side. We now need to find out an expression similar to Eq.~\ref{eq:green_equality} for the acoustic waves. \\

\noindent The propagator can be found by taking the matrix element of the evolution operator $U(t, 0)$ between the final and initial states $\ket{q}$ and $\ket{q'}$,  (see e. g. \citep{cohen_mecanique}) :

\begin{equation}
\label{eq:quantum_propagator}
K(q, q', t) = \langle q|U(t, 0)|q' \rangle.
\end{equation}

\noindent $U(t,0)$ satisfies the wave equation : $\partial^2 / \partial t^2  U(t) = -\hat{H}^2 U(t,0)$, where $\hat{H}$ is the Hamiltonian operator with $\hat{H}^2 = c_s^2 \nabla^2 + \omega_c^2 $. Then:

\begin{equation}
U(t,0) = P_1 e^{-i \hat{H} t} + P_2 e^{i \hat{H} t},
\end{equation}

\noindent with $P_1$ and $P_2$ projectors in two subspaces as above. Let ${\ket{\phi_n}}$ be the eigenfunctions of the Hamiltonian $\hat{H}$. Then, from the closure relation, we have :

\begin{align}
U(t,0) = \sum_n \ket{\phi_n} \bra{\phi_n} \left( A_1 e^{-i \omega_n t} + A_2 e^{i \omega_n t} \right), \label{eq:evolution_operator}
\end{align}

\noindent where $A_1$ and $A_2$ are the projection of the intial wave function on the two subspaces. Using Eq.~\ref{eq:evolution_operator} in Eq.~\ref{eq:quantum_propagator} gives :

\begin{equation}
K(q, q', t) = \sum_n \phi_n^*(q)\phi_n(q') \, \left( A_1e^{-i \omega_n t} + A_2 e^{i \omega_n t} \right),
\end{equation}

\noindent The $\omega$ dependent Green's function $G(q'', q', \omega)$ stems from the Fourier transform of $K(q, q', 0)$: %We write it as a sum of two terms $G = 1/2 \, (G_{+} + G_{-})$

\begin{equation}
\label{eq:greenex}
\begin{aligned}
G_{\pm}(q, q', \omega) &= \sum_n \phi_n^*(q)\phi_n(q') \frac{1}{i} \int_0^\infty \, e^{-i(\omega \pm \omega_n)t \red+ \epsilon t} \R{d}t \\
&=A_1 \sum_n \frac{\phi_n^*(q)\phi_n(q')}{\omega-\omega_n+i\epsilon} + A_2 \sum_n \frac{\phi_n^*(q)\phi_n(q')}{\omega+\omega_n+i\epsilon}.
\end{aligned}
\end{equation}

%\noindent which are computed using the property of the step function : $\int \theta(t) f'(t) \R{d}t = - f(0)$, where the prime denotes the derivative. With $\tilde{f}(t) = e^{i(\omega \pm \omega_n)t}$ we have $I = \int_0^\infty \theta(t) \tilde{f}'(t) = - \tilde{f}(0) = -1$.

%\begin{align}
%\frac{1}{i} \int_0^\infty \theta(t) \, e^{i(\omega \pm \omega_n)t} \R{d}t &= %\frac{1}{i} \times \left[ \left(i (\omega \pm \omega_n)\right)^{-1} I \right] \\
%&= \frac{1}{\omega \pm \omega_n}
%\end{align}

%\begin{equation}
%\label{eq:greenex}
%G(q'', q', \omega) = \frac{1}{2}\left( \sum_n %\frac{\phi_n^*(q)\phi_n(q')}{\omega-\omega_n} + \sum_n %\frac{\phi_n^*(q'')\phi_n(q')}{\omega+\omega_n} \right)
%\end{equation}

On the other hand, the semiclassical Green's function is obtained by taking the Fourier transform of the semiclassical propagator \ref{eq:sc_propagator} and evaluating it by stationary phase; one starts from

\begin{equation}
\label{eq:green_tf}
G_{sc}(q, q', \omega) = \frac{1}{i} \int_0^\infty K_{sc}(q, q', t) e^{i \omega t} \R{d}t.
\end{equation}

\noindent The phase term of the integrand is the action $R(q, q', t) + \omega t = S(q, q', \omega)$ or $-R(q, q', t) + \omega t = S'(q, q', \omega)$. %The function $G_{sc}(q, q', \omega)$ depends only on $R$ and $S$ trough the amplitude and phase respectively. We new need to find a sensible expression for $G_{sc}(q'', q', \omega)$. This is done in \cite{gutzwiller_chaos_1990} by 
Stationary points of the first sum are such that $\frac{\partial R(q, q', t)}{\partial t} + \omega =0$ which correspond to classical trajectories from $q'$ to $q$ at frequency $\omega$. As usual one expands the integrand in Eq.~\ref{eq:green_tf} in powers of $t$ at second order. Then the integral is approximated by the method of stationary phase \citep{schulman_techniques_1996}. Again, the computation of the prefactor requires to combine the prefactor of \ref{eq:sc_propagator} with the one coming from the stationary phase. 
The second sum has stationary points at or $\frac{\partial R(q, q', t)}{\partial t} - \omega =0$.
The result is:

\begin{equation}
\label{eq:Greensc}
\begin{aligned}
G_{sc}(q ,q',\omega) & = A_1\frac{1}{i\sqrt{2\pi}}\sum_j  \left| \frac{1}{\dot{q}\dot{q}'}\frac{\partial^2 S}{\partial q_{\perp} \partial q_{\perp}'} \right|^{1/2} e^{i S_j(q,q',\omega) - i \kappa'_j \pi/2}\\
& +A_2\frac{1}{i\sqrt{2\pi}}\sum_j  \left| \frac{1}{\dot{q}\dot{q}'}\frac{\partial^2 S}{\partial q_{\perp} \partial q_{\perp}'} \right|^{1/2} e^{i S'_j(q,q',\omega) - i \kappa'_j \pi/2},
\end{aligned}
\end{equation}

\noindent where the sum is over all classical trajectories from $q'$ to $q$ at frequency $\omega$, $\dot{q}$ and $\dot{q}'$ are final and initial velocities, 
and $q_{\perp}$ and  $q_{\perp}'$ are coordinates transverse to the orbit. The index $\kappa'_j$ counts again the singularities along the orbits.

\subsection{The final formula}

To obtain the trace formula, we should compute the trace of the Green's function from the two formulas we obtained, Eq.~Eq.~\ref{eq:greenex} and Eq.~\ref{eq:Greensc}. We will now keep only the first part in both equations since each term in one equation is equal to its counterpart in the other. Let us first compute the trace of Green's function from Eq.~\ref{eq:greenex}. To this aim, let the small imaginary part in the denominator go to zero and get for the imaginary part:

%$G(q, q, \omega) = \int G(q'',q',\omega) \R{d}^N q = \sum_j \frac{1}{\omega - \omega_j}$. This trace is related to the spectral density $\sum_j \delta(\omega - \omega_j)$. Indeed, by adding

%\begin{equation}
%\mbox{Tr} G(q', q, \omega) = \lim_{\epsilon \to 0} \frac{1}{\omega - \omega_j + i \epsilon}
%\end{equation}

%\noindent we get the spectral density as

\begin{equation}
\label{eq:Green1}
\R{Im} \; \mbox{Tr} \; G(q, q', \omega) = -\frac{1}{\pi} \, \sum_j \delta(\omega - \omega_j).
\end{equation}

\noindent On the other hand, the trace of the semiclassical Green's function Eq.~\ref{eq:Greensc} is:

\begin{equation}
\label{eq:Greentr}
\begin{aligned}
\mbox{Tr} \; G_{sc} & = \int dq \; G_{sc}(q ,q,\omega) \\
      &= \int dq   \frac{1}{i\sqrt{2\pi}}\sum_j  \left| \frac{1}{\dot{q}\dot{q}'}\frac{\partial^2 S}{\partial q_{\perp} \partial q_{\perp}'} \right|^{1/2} e^{i S_j(q,q',\omega) - i \kappa'_j \pi/2},
\end{aligned}
\end{equation}

This formula is an integral involving all closed classical paths from $q$ to $q$ at frequency $\omega$. It contains two parts. The first one corresponds to the limit for $q\rightarrow q'$ of the short direct trajectories between $q$ and $q'$, which become of zero length. We call it $\mbox{Tr} \; G_{0}$ and it should be treated separately. The remaining contains a sum of closed orbits between $q$ and $q$ with nonzero length. We evaluate this sum again by stationary phase; the stationary points in the sum are such that the first derivative of the function in the exponential is cancelled. 
This implies that $\frac{\partial S (q,q',\omega)}{\partial q'} + \frac{\partial S(q,q',\omega)}{\partial q } = 0  $ for $q=q'$. This selects closed trajectories with equal initial and final momentum, thus periodic orbits. Again, the prefactors are to be combined correctly, yielding to:

\begin{equation}
\label{eq:Greentrace}
\mbox{Tr} \; G_{sc}(q ,q',\omega) = \mbox{Tr} \; G_{0} + \sum_j  \frac{1}{i} T_j \left|\det (M_i-I)  \right|^{-1/2} e^{i S_j(\omega) - i \kappa'_j \pi/2},
\end{equation}

where $i$ labels all the periodic orbits of the system, including repetitions of a primitive orbit, $M_i$ is the monodromy matrix describing the linearized motion in the transverse direction to the orbit; the determinant encodes the stability of this orbit $i$. $T_i$ is the geometrical period of the orbit (i.e. without counting the repetitions).

Eq.~\ref{eq:Green1} connects the trace of the Green's function to the density of states $\sum_j \delta(\omega - \omega_j)$. It is known that this sum can be split in two parts $d(\omega)=\bar{d}(\omega)+d^{osc}(\omega)$. The first term corresponds to the smooth part of the density of states, while the second part contains the fluctuating (oscillatory) part. It turns out that $\bar{d}(\omega)$ corresponds to $\mbox{Tr} \; G_{0}$, while the oscillatory part corresponds to the remaining part of Eq.~\ref{eq:Greentrace}. Putting together Eq.~\ref{eq:Green1} for positive frequencies and Eq.~\ref{eq:Greentrace} gives the Gutzwiller formula for the acoustic waves:

\begin{equation}
\label{eq:Gutz}
d^{osc} (\omega) = \frac{1}{\pi} \mbox {Re} \;\sum_j   T_j \left|\det (M_i-I)  \right|^{-1/2} e^{i S_j(\omega) - i \kappa'_j \pi/2}.
\end{equation}

\bibliographystyle{aa}
\bibliography{biblio.bib}

\begin{thebibliography}{45}
\expandafter\ifx\csname natexlab\endcsname\relax\def\natexlab#1{#1}\fi

\bibitem[{Aerts {et~al.}(2010)Aerts, Christensen-Dalsgaard, \&
  Kurtz}]{aerts_asteroseismology_2010}
Aerts, C., Christensen-Dalsgaard, J., \& Kurtz, D.~W. 2010, Asteroseismology,
  Astronomy \& {Astrophysics} {Library} (Springer Netherlands)

\bibitem[{{Ballot} {et~al.}(2013){Ballot}, {Ligni{\`e}res}, \&
  {Reese}}]{Ballot2013}
{Ballot}, J., {Ligni{\`e}res}, F., \& {Reese}, D.~R. 2013, {Numerical
  Exploration of Oscillation Modes in Rapidly Rotating Stars}, ed. M.~{Goupil},
  K.~{Belkacem}, C.~{Neiner}, F.~{Ligni{\`e}res}, \& J.~J. {Green}, Vol. 865,
  91

\bibitem[{Berry(1985)}]{berry_semiclassical_1985}
Berry, M.~V. 1985, Proc. R. Soc. Lond. A, 400, 229

\bibitem[{Berry \& Robnik(1984)}]{berry_semiclassical_1984}
Berry, M.~V. \& Robnik, M. 1984, Journal of Physics A: Mathematical and
  General, 17, 2413

\bibitem[{Bogomolny \& Hugues(1998)}]{Hugues_98}
Bogomolny, E. \& Hugues, E. 1998, Physical Review E, 57, 5404

\bibitem[{Bogomolny {et~al.}(1992)Bogomolny, Georgeot, Giannoni, \&
  Schmit}]{arithm_1992}
Bogomolny, E.~B., Georgeot, B., Giannoni, M.-J., \& Schmit, C. 1992, Physical
  Review Letters, 69, 1477

\bibitem[{Bogomolny {et~al.}(1997)Bogomolny, Georgeot, Giannoni, \&
  Schmit}]{arithm_1997}
Bogomolny, E.~B., Georgeot, B., Giannoni, M.-J., \& Schmit, C. 1997, Physics
  Reports, 291, 219

\bibitem[{Bogomolny \& Keating(1996)}]{bogomolny_gutzwillers_1996}
Bogomolny, E.~B. \& Keating, J.~P. 1996, Physical Review Letters, 77, 1472

\bibitem[{Bohigas(1991)}]{bohigas_houches_1989}
Bohigas, O. 1991, Random matrix theories and chaotic dynamics, M.-J. Giannoni,
  A. Voros, and J. Zinn-Justin, Proceedings of the Les Houches Summer School of
  Theoretical Physics, LII (North-Holland, Amsterdam), 87--199

\bibitem[{Bohigas {et~al.}(1993)Bohigas, Tomsovic, \&
  Ullmo}]{bohigas_manifestations_1993}
Bohigas, O., Tomsovic, S., \& Ullmo, D. 1993, Physics Reports, 223, 43

\bibitem[{{Bowman} \& {Kurtz}(2018)}]{Bowman2018}
{Bowman}, D.~M. \& {Kurtz}, D.~W. 2018, Monthly Notices of the Royal
  Astronomical Society, 476, 3169

\bibitem[{Chang \& Shi(1986)}]{chang_evolution_1986}
Chang, S.-J. \& Shi, K.-J. 1986, Physical Review A, 34, 7

\bibitem[{Cohen-Tannoudji {et~al.}(1977)Cohen-Tannoudji, Diu, \&
  Laloe}]{cohen_mecanique}
Cohen-Tannoudji, C., Diu, B., \& Laloe, F. 1977, Mecanique quantique (293 rue
  Lecourbe, 75015 Paris: Hermann)

\bibitem[{Cvitanovic {et~al.}(2017)Cvitanovic, Artuso, Mainieri, Tanner, \&
  Vattay}]{chaosbook}
Cvitanovic, P., Artuso, R., Mainieri, R., Tanner, G., \& Vattay, G. 2017,
  Chaos: Classical and Quantum (http://chaosbook.org/)

\bibitem[{Evano {et~al.}(2019)Evano, Georgeot, \&
  Ligni\`eres}]{evano_correlations_2019}
Evano, B., Georgeot, B., \& Ligni\`eres, F. 2019, EPL, 125, 49002

\bibitem[{Garc\'ia~Hern\'andez {et~al.}(2015)Garc\'ia~Hern\'andez,
  Mart\'in-Ruiz, Monteiro, Su\'arez, Reese, Pascual-Granado, \&
  Garrido}]{garcia_hernandez_observational_2015}
Garc\'ia~Hern\'andez, A., Mart\'in-Ruiz, S., Monteiro, M. J. P. F.~G., {et~al.}
  2015, The Astrophysical Journal Letters, 811, L29

\bibitem[{Garc\'ia~Hern\'andez {et~al.}(2009)Garc\'ia~Hern\'andez, Moya,
  Michel, Garrido, Su\'arez, Rodr\'iguez, Amado, Mart\'in-Ruiz, Rolland,
  Poretti, Samadi, Baglin, Auvergne, Catala, Lefevre, \&
  Baudin}]{hernandez_asteroseismic_2009}
Garc\'ia~Hern\'andez, A., Moya, A., Michel, E., {et~al.} 2009, Astronomy \&
  Astrophysics, 506, 79

\bibitem[{{Garc\'ia Hern\'andez} {et~al.}(2013){Garc\'ia Hern\'andez}, {Moya},
  {Michel}, {Su\'arez}, {Poretti}, {Mart\'in-Ru\'iz}, {Amado}, {Garrido},
  {Rodr\'iguez}, \& {Rainer}}]{Hernandez2013}
{Garc\'ia Hern\'andez}, A., {Moya}, A., {Michel}, E., {et~al.} 2013, Astronomy
  \& Astrophysics, 559, A63

\bibitem[{Gough(1993)}]{gough_linear_1993}
Gough, D. 1993, Linear adiabatic stellar pulsation, J-P. Zahn and J.
  Zinn-Justin, Proceedings of the Les Houches Summer School of Theoretical
  Physics, XLVII (Elsevier, Amsterdam), 400--560

\bibitem[{Gutzwiller(1990)}]{gutzwiller_chaos_1990}
Gutzwiller, M.~C. 1990, Chaos in {Classical} and {Quantum} {Mechanics},
  Interdisciplinary {Applied} {Mathematics} (New York: Springer-Verlag)

\bibitem[{Hannay \& Ozorio De~Almeida(1984)}]{Hannay_84}
Hannay, J.~H. \& Ozorio De~Almeida, A.~M. 1984, Journal of Physics A:
  Mathematical and General, 17, 3429

\bibitem[{Hansen {et~al.}(2004)Hansen, Kawaler, \&
  Trimble}]{hansen_stellar_2004}
Hansen, C.~J., Kawaler, S.~D., \& Trimble, V. 2004, Stellar {Interiors}:
  {Physical} {Principles}, {Structure}, and {Evolution}, 2nd edn., Astronomy \&
  {Astrophysics} {Library} (New York: Springer-Verlag)

\bibitem[{Kudrolli {et~al.}(1994)Kudrolli, Sridhar, Pandey, \&
  Ramaswamy}]{kudrolli_signatures_1994}
Kudrolli, A., Sridhar, S., Pandey, A., \& Ramaswamy, R. 1994, Physical Review
  E, 49, R11

\bibitem[{Ligni\`eres \& Georgeot(2008)}]{lignieres_wave_2008}
Ligni\`eres, F. \& Georgeot, B. 2008, Physical Review E, 78, 016215

\bibitem[{Ligni\`eres \& Georgeot(2009)}]{lignieres_asymptotic_2009}
Ligni\`eres, F. \& Georgeot, B. 2009, Astronomy \& Astrophysics, 500, 1173

\bibitem[{Ligni\`eres {et~al.}(2006)Ligni\`eres, Rieutord, \&
  Reese}]{lignieres_acoustic_2006}
Ligni\`eres, F., Rieutord, M., \& Reese, D. 2006, Astronomy \& Astrophysics,
  455, 607

\bibitem[{Mehta(2004)}]{metha_random_2004}
Mehta, M. 2004, Random {Matrices} (Elsevier)

\bibitem[{{Michel} {et~al.}(2017){Michel}, {Dupret}, {Reese}, {Ouazzani},
  {Debosscher}, {Hern{\'a}ndez}, {Belkacem}, {Samadi}, {Salmon}, \&
  {Suarez}}]{Michel2017}
{Michel}, E., {Dupret}, M.-A., {Reese}, D., {et~al.} 2017, in European Physical
  Journal Web of Conferences, Vol. 160, 03001

\bibitem[{Mirouh {et~al.}(2019)Mirouh, Angelou, Reese, \&
  Costa}]{mirouh_mode_2019}
Mirouh, G.~M., Angelou, G.~C., Reese, D.~R., \& Costa, G. 2019, Monthly Notices
  of the Royal Astronomical Society: Letters, 483, L28

\bibitem[{Ott(1993)}]{ott_chaos_1993}
Ott, E. 1993, Chaos in {Dynamical} {Systems} (Cambridge University Press)

\bibitem[{Ouazzani {et~al.}(2012)Ouazzani, Dupret, \&
  Reese}]{ouazzani_pulsations_2012}
Ouazzani, R.-M., Dupret, M.-A., \& Reese, D.~R. 2012, Astronomy \&
  Astrophysics, 547, A75

\bibitem[{Ouazzani {et~al.}(2015)Ouazzani, Roxburgh, \&
  Dupret}]{ouazzani_pulsations_2015}
Ouazzani, R.-M., Roxburgh, I.~W., \& Dupret, M.-A. 2015, Astronomy \&
  Astrophysics, 579, A116

\bibitem[{O’Connor {et~al.}(1987)O’Connor, Gehlen, \&
  Heller}]{oconnor_properties_1987}
O’Connor, P., Gehlen, J., \& Heller, E.~J. 1987, Physical Review Letters, 58,
  1296

\bibitem[{{Papar\'o} {et~al.}(2016){Papar\'o}, {Benk\"o}, {Hareter}, \&
  {Guzik}}]{Paparo2016}
{Papar\'o}, M., {Benk\"o}, J.~M., {Hareter}, M., \& {Guzik}, J.~A. 2016,
  Astrophysical Journal Supplement Series, 224, 41

\bibitem[{Pasek {et~al.}(2011)Pasek, Georgeot, Ligni\`eres, \&
  Reese}]{pasek_regular_2011}
Pasek, M., Georgeot, B., Ligni\`eres, F., \& Reese, D.~R. 2011, Physical Review
  Letters, 107, 121101

\bibitem[{Pasek {et~al.}(2012)Pasek, Ligni{\`e}res, Georgeot, \&
  Reese}]{pasek_regular_2012}
Pasek, M., Ligni{\`e}res, F., Georgeot, B., \& Reese, D.~R. 2012, Astronomy \&
  Astrophysics, 546, A11

\bibitem[{Reese {et~al.}(2006)Reese, Ligni\`eres, \&
  Rieutord}]{reese_acoustic_2006}
Reese, D., Ligni\`eres, F., \& Rieutord, M. 2006, Astronomy \& Astrophysics,
  455, 621

\bibitem[{Reese {et~al.}(2008)Reese, Ligni\`eres, \&
  Rieutord}]{reese_regular_2008}
Reese, D., Ligni\`eres, F., \& Rieutord, M. 2008, Astronomy \& Astrophysics,
  481, 449

\bibitem[{Reese {et~al.}(2017)Reese, Ligni\`eres, Ballot, Dupret, Barban,
  Veer-Menneret, \& MacGregor}]{reese_frequency_2017}
Reese, D.~R., Ligni\`eres, F., Ballot, J., {et~al.} 2017, Astronomy \&
  Astrophysics, 601, A130

\bibitem[{Reese {et~al.}(2009)Reese, MacGregor, Jackson, Skumanich, \&
  Metcalfe}]{reese_pulsation_2009}
Reese, D.~R., MacGregor, K.~B., Jackson, S., Skumanich, A., \& Metcalfe, T.~S.
  2009, Astronomy \& Astrophysics, 506, 189

\bibitem[{Reese {et~al.}(2013)Reese, Prat, Barban, Veer-Menneret, \&
  MacGregor}]{reese_mode_2013}
Reese, D.~R., Prat, V., Barban, C., Veer-Menneret, C. v., \& MacGregor,
  K.~B. 2013, Astronomy \& Astrophysics, 550, A77

\bibitem[{Schulman(1996)}]{schulman_techniques_1996}
Schulman, L.~S. 1996, {Techniques} and {Applications} of {Path} {Integration}
  (Wiley Classics Library)

\bibitem[{Shim {et~al.}(2011)Shim, Wiersig, \& Cao}]{shim_whispering_2011}
Shim, J.-B., Wiersig, J., \& Cao, H. 2011, Physical Review E, 84, 035202

\bibitem[{Sieber \& Richter(2001)}]{sieber_correlations_2001}
Sieber, M. \& Richter, K. 2001, Physica Scripta, 2001, 128

\bibitem[{Vidmar {et~al.}(2007)Vidmar, St\"ockmann, Robnik, Kuhl, H\"ohmann, \&
  Grossmann}]{vidmar2007beyond}
Vidmar, G., St\"ockmann, H.-J., Robnik, M., {et~al.} 2007, Journal of Physics
  A: Mathematical and Theoretical, 40, 13883

\end{thebibliography}

\end{document}